\documentstyle[prd,aps,,graphics,psfrag,epsf,epsfig,psfig]{revtex}
\begin{document}
\title{
\hfill{\small {FZJ-IKP(TH)-1999-10}}\\
\hfill{\small {MKPH-98-11}}\\[2cm]
Generalized Polarizabilities of the Nucleon in Chiral Effective Theories}

\vspace{1cm}

\author{Thomas R. Hemmert$^a$\footnote{email: th.hemmert@fz-juelich.de},
Barry R. Holstein$^{a,b}$\footnote{email: holstein@phast.umass.edu}
Germar Kn{\" o}chlein$^c$\footnote{email: knoechle@kph.uni-mainz.de}, and
Dieter Drechsel$^c$\footnote{email: drechsel@kph.uni-mainz.de}}

\vspace{0.5cm}

\address{$^a$ Forschungszentrum J\"ulich, Institut f\"ur Kernphysik (Th),
D-52425 J\"ulich, Germany }
\address{$^b$ Department of Physics and Astronomy, University of Massachusetts,
Amherst, MA 01003, USA}
\address{$^c$ Institut f{\" u}r Kernphysik, Johannes Gutenberg-Universit{\"
a}t, D-55099 Mainz, Germany}
\maketitle

\thispagestyle{empty}

\vspace{2cm}

\begin{abstract}
Using the techniques of chiral effective field theories we evaluate the
so called ``generalized polarizabilities", which characterize the
structure dependent components in virtual Compton scattering (VCS) off the nucleon
as probed in the electron scattering reaction $eN\rightarrow e'N\gamma$.
Results are given for both spin-dependent and spin-independent structure
effects to ${\cal O}(p^3)$ in SU(2) Heavy Baryon Chiral Perturbation Theory
and
to ${\cal O}(\epsilon^3)$ in the SU(2) ``Small Scale Expansion''. 
\end{abstract}
\newpage

\section{Introduction}
One of the primary goals of contemporary particle/nuclear physics is to
understand
the structure of the nucleon.  Indeed this is being pursued at the very highest
energy
machines such as HERA and SLAC, at which one probes the quark/parton
substructure,
as well as at lower energy accelerators such as BATES, ELSA and MAMI, wherein one
studies the
low energy structure of the nucleon via electron scattering. In recent
years another important low energy probe has been (real) Compton scattering, 
by which one
can study the deformation of the nucleon under the influence of quasi-static
electric
and/or magnetic fields \cite{LV}. For example, in the presence of an external
electric field
$\vec{E}$ the quark distribution of the nucleon becomes distorted, leading
to an induced electric dipole moment
\begin{equation}
\vec{p}=4\pi\alpha_E\vec{E}
\end{equation}
in the direction of the applied field, where $\alpha_E$ is the electric
polarizability.
The interaction of this dipole moment with the field leads to a
corresponding interaction energy
\begin{equation}
U=-{1\over 2}4\pi\alpha_E\vec{E}^2.
\end{equation}
Similarly in the presence of an applied magnetizing field $\vec{H}$ there will
be
an induced magnetic dipole moment
\begin{equation}
\vec{\mu}=4\pi\beta_M\vec{H}
\end{equation}
and an interaction energy
\begin{equation}
U=-{1\over 2}4\pi\beta_M\vec{H}^2.
\end{equation}
For wavelengths large compared to the size of the system, the effective
Hamiltonian
for the interaction of a system of charge $e$ and mass $m$ with an
electromagnetic
field is, of course, given by the simple form
\begin{equation}
H^{(0)}={(\vec{p}-e\vec{A})^2\over 2m}+e\phi,
\end{equation}
and the Compton scattering cross section has simply the familiar Thomson form
\begin{equation}
{d\sigma\over d\Omega}=\left({\alpha_{em}\over m}\right)^2\left({\omega'\over
\omega}\right)^2[{1\over 2}
(1+\cos^2\theta)],
\end{equation}
where $\alpha_{em}$ is the fine structure constant and $\omega,\omega'$ are the
initial, final photon energies respectively.
As the energy increases, however, so does the resolution and
one must take into account also polarizability
effects, whereby the effective Hamiltonian becomes
\begin{equation}
H_{\rm eff}=H^{(0)}-{1\over 2}4\pi(\alpha_E\vec{E}^2+\beta_M\vec{H}^2).
\end{equation}
The Compton scattering cross section from such a system (taken, for simplicity,
to be spinless) is given then by
\begin{eqnarray}
{d\sigma\over d\Omega}&=&\left({\alpha_{em}\over m}\right)^2\left({\omega'\over
\omega}\right)^2[{1\over 2}
(1+\cos^2\theta)\nonumber\\
&-&{m\omega\omega'\over \alpha_{em}}[{1\over
2}(\alpha_E+\beta_M)(1+\cos\theta)^2
+{1\over 2}(\alpha_E-\beta_M)(1-\cos\theta)^2+\ldots].\label{eq:sss}
\end{eqnarray}
It is clear from Eq.(\ref{eq:sss})
that from careful measurement of the differential scattering cross section,
extraction of these structure dependent polarizability terms is possible
provided that
i) the energy is large enough that these terms are significant compared to the
leading
Thomson piece and ii) that the energy is not so large that higher order
corrections become important.  In this way the measurement of electric and
magnetic polarizabilities for the proton has recently been accomplished using
photons in
the energy range 50 MeV  $<\omega <$ 100 MeV, yielding\cite{PPol}
\footnote{Results for the neutron extracted from $n-Pb$ scattering cross
section
measurements have been reported\cite{npol} but have been questioned\cite{ques}.
Extraction via studies using a deuterium target may be possible
in the future\cite{bean}.}
\begin{eqnarray}
\alpha_E^p&=&(12.1\pm 0.8\pm 0.5)\times 10^{-4}\; {\rm fm}^3\nonumber\\
\beta_M^p&=&(2.1\mp 0.8\mp 0.5)\times 10^{-4}\; {\rm fm}^3. \label{abexp}
\end{eqnarray}
Note that in
practice one generally exploits the strictures of causality and unitarity as
manifested
in the validity of the forward scattering dispersion relation, which yields the
Baldin sum rule\cite{bgm}
\begin{equation}
\alpha_E^{p,n}+\beta_M^{p,n}={1\over 2\pi^2}\int_0^\infty{d\omega\over
\omega^2}
\sigma_{\rm tot}^{p,n}=\left\{
\begin{array}{ll}(13.69\pm 0.14)\times 10^{-4}{\rm fm}^3& {\rm proton}\\
                 (14.40\pm 0.66)\times 10^{-4}{\rm fm}^3& {\rm neutron}
\end{array}\right.
\end{equation}
as a rather precise constraint because of the small uncertainty associated 
with the photoabsorption cross section $\sigma_{\rm tot}^p$.

{From} these results, which imply that the polarizabilities of the proton are
nearly a factor of a thousand smaller than its volume, we learn
that the nucleon is a relatively rigid object when compared to the
hydrogen atom, for example, for which the electric polarizability and volume
are comparable.

Additional probes of proton structure are possible if one exploits its
spin $\vec{S}$.  Thus, for example, the presence of a time varying electric field
in the plane
of a rotating system of charges will lead to a charge separation with induced
electric dipole moment
\begin{equation}
\vec{p}=-\gamma_1\vec{S}\times{\partial \vec{E}\over \partial t}
\end{equation}
and corresponding interaction energy
\begin{equation}
U_1=-\vec{p}\cdot\vec{E}=\gamma_1\vec{E}\cdot\vec{S}\times
(\vec{\nabla}\times\vec{B}),\label{spinpola1}
\end{equation}
where we have used the Maxwell equations in writing this form.  This is a
quantum mechanical analog of the familiar Faraday rotation.  (Note that the
"extra" time or spatial derivative is required by time reversal invariance
since
$\vec{S}$ is T-odd.)  Similarly other possible structures are\cite{Ragusa,NS}
\begin{eqnarray}\label{spinpolas}
U_2&=&\gamma_2\vec{B}\cdot\vec{\nabla}\vec{S}\cdot\vec{E}\nonumber\\
U_3&=&\gamma_3\vec{E}\cdot\vec{\nabla}\vec{S}\cdot\vec{B}\nonumber\\
U_4&=&\gamma_4\vec{B}\cdot\vec{S}\times(\vec{\nabla}\times\vec{E}),
\end{eqnarray}
and the measurement of these various "spin-polarizabilities" $\gamma_i$ via
polarized Compton scattering provides a rather different probe for
nucleon
structure. Because of the requirement for polarization not much is known at
present about
such spin-polarizabilities, although from dispersion relations the
combination
\begin{equation}
\gamma_0^p\equiv \gamma_1^p-\gamma_2^p-2\gamma_4^p\approx\left\{ 
\begin{array}{cc}
-1.34\times 10^{-4}\; {\rm fm}^4 & {\rm SAID}\cite{DHS}\\
-0.80\times 10^{-4}\; {\rm fm}^4 & {\rm Mainz}\cite{HDT}
\end{array}\right.
\end{equation}
has been evaluated and from a global analysis of unpolarized Compton data, to
which it contributes at ${\cal O}(\omega^4)$, Tonnison et al.\cite{ton} have 
determined the so-called backward spin-polarizability to be
\begin{equation}
\gamma_\pi=\gamma_1+\gamma_2+2\gamma_4=(27.7\pm 2.3\pm 2.5)\times 10^{-4}\;
{\rm
fm}^4.
\end{equation}
Clearly such measurements represent an important goal for the future.

At the same time it has come to be realized that a high
resolution probe of nucleon structure is available, in principle,
via the use of {\it virtual} Compton
scattering---VCS---wherein virtual photons produced from scattered electrons
are scattered off a nucleon into real final state photons, transferring a 
three-momentum $\bar{q}$ to the target.  The outcome of
such measurements is, in principle, $\bar{q}$-dependent 
values of the polarizabilities
(usually termed "generalized polarizabilities" and denoted by GPs in the following) 
which can be thought of as the Fourier transforms of
{\it local} polarization densities in the nucleon.  At the present time a VCS
experiment has already taken place at MAMI, and there
exist approved experiments at BATES and TJNAF.  Preliminary
results have been reported from MAMI and will be discussed in the
conclusion \cite{nstar}.  It
is therefore appropriate to have a base of solid theoretical predictions with
which such
data can be confronted. The here presented approach, which utilizes the 
techniques of chiral effective theories in the heavy fermion formulation, 
has already yielded several results\cite{HHKS1,HHKS2}.  In
the first chiral calculation of generalized polarizabilities
utilizing SU(2) Heavy Baryon Chiral Perturbation Theory (HBChPT) \cite{HHKS1},
the leading momentum-dependent modification of
the (generalized) electric ($\bar{\alpha}_E(\bar{q})$) and magnetic
($\bar{\beta}_M(\bar{q})$) polarizabilities was analyzed.
Later, in a short communication\cite{HHKS2}, 
numerical studies for the full $\bar{q}$-dependence of
all 10 generalized (Guichon) polarizabilities were presented---again using
the framework of SU(2) HBChPT.
In this work we present the details
behind the numerical study of ref.\cite{HHKS2} and, for the first time in the field of
VCS,
are able to present simple analytical expressions for all GPs in a momentum
range
from $0<\bar{q}^2<0.5$ GeV$^2$ utilizing SU(2) HBChPT. 
These new expressions greatly facilitate the study of the 
influence of the chiral ``pion cloud'' on the GPs and the comparison with model
calculations. Furthermore, we also investigate the leading
modifications
of the GPs' $\bar{q}$-dependence due to $\Delta$(1232) resonance contributions
utilizing a different effective chiral lagrangian approach---the so called
``small scale expansion'' (SSE) \cite{HBDel}.
SSE results have already been reported for real 
Compton scattering \cite{delta,HHKK}, 
and in the present work we generalize the analysis to the VCS case.

In the next section we shall discuss the definition of the
generalized polarizabilities, while in Section III we present an introduction
to
the way in which our heavy baryon calculations---valid to one loop---are
carried out. In section IV we show how to connect our predictions to the
general formulation of VCS and how to extract the desired generalized
polarizabilities.
In section V, we present the results of our calculations.
Finally, we summarize our findings in a concluding
section VI.

\section{Generalized Polarizabilities}
Recently a new frontier in Compton scattering has been opened (see, {\it e.g.},
\cite{NF}) and is now in the beginning of being explored:
the study of the  electron scattering process
$e p \rightarrow e' p' \gamma$ ({\it cf. Fig. 1}) in order
to obtain information concerning the virtual Compton scattering\footnote{Chiral
analyses of double virtual Compton scattering $\gamma^\ast p\rightarrow \gamma^\ast p$
in the forward direction and its connection with the spin structure of the nucleon
have recently been published \cite{dvcs}.} (VCS) process
$\gamma^* N \rightarrow \gamma N$.
As will be discussed below, in addition to the two kinematical variables
of real Compton scattering---\ the scattering angle $\theta$ and the
energy $\omega'$ of the outgoing photon---the invariant structure functions
for VCS \cite{BAS,Guichon} depend on a {\em third} kinematical
variable, {\it e.g.} the magnitude of the three--momentum transfer to the nucleon in the
hadronic c.m. frame, $\bar{q}\equiv|\vec{q}|$. As shown in ref. \cite{Guichon},
the VCS amplitude can then be characterized in terms of
$\bar{q}$-dependent GPs, in analogy to the
well-known polarizability coefficients in real Compton
scattering. However, due to the specific kinematic approximation chosen in
\cite{Guichon}
there does not exist a one--to--one correspondence between
the real Compton polarizabilities and the GPs of Guichon et al.
in VCS\cite{Guichon,fm1,fm2}.

The advantage of VCS lies in the virtual nature of the
initial state photon and the associated possibility of
an {\it independent} variation of photon energy and momentum,
thus rendering access to a much greater variety of structure
information than in the case of real Compton scattering.
For example, one can hope to identify individual signatures of specific
nucleon resonances which cannot be obtained in other
processes\cite{NF}.  In this regard,
it should be noted that a great deal of theoretical work
already exists, such as predictions  within a non--relativistic 
constituent quark
model\cite{Guichon}, a one--loop calculation in the linear sigma
model\cite{Metz}, a Born term model including
nucleon resonance effects\cite{Vanderhaeghen}, a HBChPT calculation of the leading
$\bar{q}$-dependence of the generalized electric and magnetic polarizability 
\cite{HHKS1}, a
calculation of $\bar{\alpha}_E(\bar{q}^2)$ in the Skyrme model\cite{KM97}
and the numerical study of all 10 GPs again utilizing HBChPT \cite{HHKS2}.
For an overview of the status at higher energies and in the
deep inelastic regime we refer to \cite{NF}.  

The GPs of the nucleon have been defined by Guichon et al.
in terms of electromagnetic multipoles
as functions of the initial photon momentum $\bar{q}$\cite{Guichon},
\begin{eqnarray}
P^{(\rho' L' , \rho L)S} (\bar{q}^2)
& = &
\left[ \frac{1}{\omega'^{L} \bar{q}^{L}}
H^{(\rho' L' , \rho L)S} (\omega' , \bar{q}) \right]_{\omega' = 0} \, ,
\nonumber\\
\hat{P}^{(\rho' L' , L)S} (\bar{q}^2) & = &
\left[ \frac{1}{\omega'^{L} \bar{q}^{L+1}}
\hat{H}^{(\rho' L' , L)S} (\omega' , \bar{q}) \right]_{\omega' = 0}
\, ,
\end{eqnarray}
\noindent where $L$ ($L'$) denotes the initial (final) photon angular
momentum,
$\rho$ ($\rho'$) the type of multipole transition ($0 = C$ (scalar, Coulomb),
$1 = M$ (magnetic),
$2 = E$ (electric)), and $S$ distinguishes between non--spin--flip ($S=0$) and
spin--flip ($S=1$) transitions.
In addition, mixed--type polarizabilities, ${\hat{P}}^{(\rho' L' , L)S}
(\bar{q}^2)$,
have been introduced, which are neither purely electric nor purely Coulomb
type.
It is important to note that the above definitions are based on the
kinematical approximation that the multipoles are expanded around $\omega' = 0$
and {\em{only terms linear in $\omega'$ are retained}}, which, together with
current
conservation, yields selection rules for the possible combinations of
quantum numbers of the GPs. In this approximation, 10 GPs
have been introduced in \cite{Guichon} as functions of $\bar{q}^2$:
$P^{(01,01)0}$,
$P^{(11,11)0}\,$,
$P^{(01,01)1}\,$,
$P^{(11,11)1}\,$,
$P^{(01,12)1}\,$,
$P^{(11,02)1}\,$,
$P^{(11,00)1}\,$,
${\hat{P}}^{(01,1)0}\,$,
${\hat{P}}^{(01,1)1}\,$,
${\hat{P}}^{(11,2)1}\,$.

However, recently it has been proven\cite{fm1,fm2}, using
crossing symmetry and charge conjugation invariance, that only
{\it six} of the above ten GPs are independent.
With
\begin{equation}
\omega_0 \equiv M_N - \sqrt{M_N^2+\bar{q}^2}=-\;\frac{\bar{q}^2}{2M_N}+
{\cal O}(1/M_N^3) \label{om0}
\end{equation}
and $M_N$ being the nucleon mass,
the four constraints implied by $C$ invariance and crossing can be written as
\begin{eqnarray}
0 & = & \sqrt{\frac{3}{2}} P^{(01,01)0}(\bar{q}^2)
 + \sqrt{\frac{3}{8}} P^{(11,11)0}(\bar{q}^2)
 + \frac{3 \bar{q}^2}{2 \omega_0} \hat{P}^{(01,1)0}(\bar{q}^2) \, ,
\nonumber\\
0 & = & P^{(11,11)1}(\bar{q}^2)
 + \sqrt{\frac{3}{2}} \omega_0 P^{(11,02)1}(\bar{q}^2)
 + \sqrt{\frac{5}{2}} \bar{q}^2 \hat{P}^{(11,2)1}(\bar{q}^2) \, ,
\nonumber\\
0 & = & 2 \omega_0 P^{(01,01)1}(\bar{q}^2)
 + 2 \frac{\bar{q}^2}{\omega_0} P^{(11,11)1}(\bar{q}^2)
 - \sqrt{2} \bar{q}^2 P^{(01,12)1}(\bar{q}^2)
 + \sqrt{6} \bar{q}^2 \hat{P}^{(01,1)1}(\bar{q}^2) \, ,
\nonumber\\
0 & = & 3 \frac{\bar{q}^2}{\omega_0} P^{(01,01)1}(\bar{q}^2)
 - \sqrt{3} P^{(11,00)1}(\bar{q}^2)
 - \sqrt{\frac{3}{2}} \bar{q}^2 P^{(11,02)1}(\bar{q}^2) \, .\label{cinv}
\end{eqnarray}
In the scalar (i.e. spin--independent) sector the first of 
Eqs.(\ref{cinv}) allows us to eliminate the mixed
polarizability ${\hat{P}}^{(01,1)0}$
in favor of $P^{(01,01)0}$ and $P^{(11,11)0}$, which are simply
generalizations of the familiar electric and magnetic polarizabilities in real
Compton scattering
\begin{eqnarray}
\bar{\alpha}_E (\bar{q}^2) & = & - \frac{e^{2}}{4 \pi} \sqrt{\frac{3}{2}}
P^{(01,01)0} (\bar{q}^2) \,,
\nonumber\\
\bar{\beta}_M (\bar{q}^2) & = & - \frac{e^{2}}{4 \pi} \sqrt{\frac{3}{8}}
P^{(11,11)0} (\bar{q}^2) \,.
\end{eqnarray}
In the limit $\bar{q}\rightarrow 0$ they reduce to the real Compton polarizabilities
$\bar{\alpha}_E,\,\bar{\beta}_M$ of Eq.(\ref{abexp}).

In the spin-dependent sector it is not a priori clear which
three of the seven GPs
$P^{(01,01)1}\,$,
$P^{(11,11)1}\,$,
$P^{(01,12)1}\,$,
$P^{(11,02)1}\,$,
$P^{(11,00)1}\,$,
${\hat{P}}^{(01,1)1}\,$,
${\hat{P}}^{(11,2)1}\,$.
should be eliminated by use of Eq.(\ref{cinv}). However, the chiral analysis 
performed here shows that to leading order only 4 of the 7 spin GPs can be 
calculated---$P^{(01,12)1},\,P^{(11,02)1},\,P^{(11,00)1},\,{\hat{P}}^{(01,1)1}$.
Naturally we focus on these four spin GPs, as 
$P^{(01,01)1},\,P^{(11,11)1},\,{\hat{P}}^{(11,2)1}$ possess an extra suppression 
factor of $1/M_N$ (see section \ref{HBChPTspin})
which pushes them outside the validitiy of our analysis. Still, one
can reconstruct the whole set of spin GPs via Eq.(\ref{cinv}) if one wishes to do so.
Finally, we note that in the spin-sector one can also establish a (partial)
connection between the GPs defined 
in the context of VCS by Guichon et al. \cite{Guichon} and the 4 real Compton 
spin polarizabilities $\gamma_i,\;i=1...4$ of Ragusa \cite{Ragusa} given in 
Eqs.(\ref{spinpola1}) and (\ref{spinpolas}):
\begin{eqnarray}\label{spinconnection}
\gamma_3&=&-\frac{e^2}{4\pi}\,\frac{3}{\sqrt{2}}\,P^{(01,12)1}(\bar{q}\rightarrow 0) 
\nonumber \\
\gamma_2+\gamma_4&=&-\frac{e^2}{4\pi}\,\frac{3\sqrt{3}}{2\sqrt{2}}\,P^{(11,02)1}
(\bar{q}\rightarrow 0)
\end{eqnarray}
These model-independent relations might provide an interesting
possibility to determine some of the elusive (Ragusa) 
spin-polarizabilities by the upcoming experiments.  

\section{The chiral framework}
\subsection{Pion-Nucleon ChPT}

We want to perform the VCS calculation to ${\cal O}(p^3)$ in Heavy Baryon
Chiral Perturbation Theory (HBChPT) ({\it e.g.} see \cite{review}).
We therefore need the lagrangians
\begin{eqnarray}
{\cal L}_{VCS}^{(3)}={\cal L}_{N}^{(3)}+{\cal L}_{\pi}^{(4)}.
\end{eqnarray}

We begin our discussion in the nucleon sector. For VCS to
${\cal O}(p^3)$ we need the lagrangians
\begin{equation}
{\cal{L}}_{N}^{(3)} = {\cal{L}}_{\pi N}^{(1)} + {\cal{L}}_{\pi N}^{(2)}
+ {\cal{L}}_{\pi N}^{(3)} \, , \label{eq:chiL}
\end{equation}
with
\begin{eqnarray}
{\cal{L}}_{\pi N}^{(1)} & = & \bar N_v(iv \cdot D + \dot{g}_A S \cdot
                              u) N_v \, , \nonumber\\
{\cal{L}}_{\pi N}^{(2)} & = &  \frac{1}{2M_0} \bar N_v \left\{(v\cdot D)^2-D^2-
                               \frac{i}{2}\left[S^\mu,S^\nu\right]\left[\left(
                               1+\dot{\kappa}_v\right)f_{\mu\nu}^+
                               +2\left(1+\dot{\kappa}_s\right)v_{\mu\nu}^{(s)}
                               \right]+\dots\right\}N_v\; , \nonumber\\
{\cal{L}}_{\pi N}^{(3)} & = & \frac{-1}{8 M_0^2} \bar N_v
                              \left\{\left(1+2\dot{\kappa}_v\right)
                              \left[S_\mu,S_\nu\right]f_+^{\mu\sigma}v_\sigma
                              D^\nu+2\left(\dot{\kappa}_s-\dot{\kappa}_v\right)
                              \left[S_\mu,S_\nu\right]v_{(s)}^{\mu\sigma}v_\sigma
D^\nu
                              + \mathrm{h.c.}+\dots\right\} N_v\; , \label{piN}
\end{eqnarray}
where we have only kept those
terms\footnote{We note that to the order we are working in the VCS calculation
the nucleon mass parameter $M_0$
can be replaced by the physical nucleon mass $M_N$, the axial-vector coupling
in the
chiral limit $\dot{g}_A$ can be replaced with the physical axial-vector
coupling constant
$g_A=1.267$ and the isoscalar [isovector] anomalous magnetic moment of the
nucleon
in the chiral limit $\dot{\kappa}_s\;[\dot{\kappa}_v]$ can be replaced with the
physical
isoscalar [isovector] anomalous magnetic moment
 $\kappa_s=\kappa_p+\kappa_n=-0.120\,
\mbox{n.m.}\;[\kappa_v=\kappa_p-\kappa_n=3.71\,\mbox{n.m.}]$. Details of the
renormalization of these parameters in the chiral lagrangian by loop effects
and
higher order counter terms can be found in ref.\cite{bfhm}, both for HBChPT and
SSE.}
which contribute to our VCS calculation. Furthermore, all terms which vanish in
the
``Coulomb gauge'' $v \cdot A=0$, with $v_\mu$ being the velocity vector
$(v^2=1)$
of the nucleon
and $A_\mu$ denoting a photon field, have been omitted.
The velocity-dependent nucleon field $N_v$ is projected from
the relativistic nucleon Dirac field $\Psi_N$ via
\begin{equation}
N_v = {\rm{exp}} \left[ i M_0 v \cdot x \right] P_v^+ \Psi_N \, ,
\end{equation}
where the velocity projection operator is given by
\begin{equation}
P_v^+ = \frac{1}{2} \left( 1 + \not\!{v} \right) \, .
\end{equation}
$S_\mu$ denotes the usual Pauli-Lubanski vector ({\it e.g.}
\cite{review})
and $D_\mu$ corresponds to the covariant derivative of the nucleon
\begin{equation}
D_{\mu}\,N_v =\left[\partial_{\mu} + \Gamma_{\mu} - i v_{\mu}^{(s)}\right]N_v.
\end{equation}
One also encounters the following chiral tensors in the VCS calculation:
\begin{eqnarray}
U & = & u^2={\mathrm{exp}} \left(i \vec \tau \cdot \vec \pi /F_\pi
\right)\nonumber \\
\Gamma_{\mu} & = & \frac{1}{2} \left\{ u^{\dagger} \left(
    \partial_{\mu} - i \,e\,\frac{\tau^3}{2}\,A_\mu\right) u
    + u \left( \partial_{\mu} - i\,e\,\frac{\tau^3}{2}\,A_\mu
    \right) u^{\dagger} \right\} \; , \nonumber\\
u_{\mu} & = & i \left\{ u^{\dagger} \left( \partial_{\mu} -
i\,e\,\frac{\tau^3}{2}\,A_\mu
    \right) u - u \left( \partial_{\mu} - i\,e\,\frac{\tau^3}{2}\,A_\mu \right)
    u^{\dagger} \right\} \; . \label{eq:aa}
\end{eqnarray}
In Eq.(\ref{eq:aa}) $\vec \tau$ are the conventional Pauli isospin
matrices, while $\vec \pi$ represents the interpolating pion field.
Furthermore, $v_\mu^{(s)}=e\,\frac{1}{2}\,A_\mu$ denotes an isoscalar photon
field
and the corresponding field strength tensors in Eq.(\ref{piN}) are defined as
\begin{eqnarray}
v_{\mu \nu}^{(s)} & = & \partial_{\mu} v_{\nu}^{(s)} - \partial_{\nu}
v_{\mu}^{(s)} \, , \nonumber\\
f_+^{\mu \nu} & = & u\,e\frac{\tau^3}{2}\,\left(\partial_\mu A_\nu-\partial_\nu
A_\mu\right) u^{\dagger} + u^{\dagger}\,e\frac{\tau^3}{2}\,\left(\partial_\mu
A_\nu-\partial_\nu A_\mu\right)  u \, .
\end{eqnarray}

{From} the pion sector we require information up to ${\cal O}(p^4)$ for a
${\cal O}(p^3)$ VCS calculation. Utilizing ``standard ChPT'' \cite{GL}
({\it
i.e.}
the assumption of a ``large'' quark condensate parameter $B$) one finds
\begin{eqnarray}
{\cal L}_{\pi}^{(4)}={\cal L}_{\pi\pi}^{(2)}+{\cal L}_{\pi\pi}^{(4)}
\end{eqnarray}
with
\begin{eqnarray}\label{Lpi}
{\cal{L}}_{\pi \pi}^{(2)} &=& \frac{F_0^2}{4} {\mathrm{tr}}
          \left[ \left(\nabla_{\mu} U\right)^{\dagger} \nabla^{\mu}
           U +\chi^\dagger U+\chi U^\dagger \right] \; , \nonumber \\
{\cal{L}}_{\pi \pi}^{(4)} &=&{e^2\over 32\pi^2 F_0}\epsilon^{\mu\nu\alpha\beta}
                             F_{\mu\nu}F_{\alpha\beta}\pi^0+\ldots \; ,
\end{eqnarray}
where again we have omitted all terms not required for the VCS calculation.
Note that the only piece shown from the chiral ${\cal O}(p^4)$ meson lagrangian
is the
so called ``anomalous'' or ``Wess-Zumino'' term \cite{WZ}, which one needs for
the
${\cal O}(p^3)$ pion-pole diagram of VCS shown in Fig.\ref{figgborn}(f). In the
lagrangians of Eq.(\ref{Lpi}) one also encounters the chiral tensors
\begin{eqnarray}
\nabla_{\mu} U &=& \partial_{\mu} U -i\,\frac{e}{2}\,A_{\mu} \left[
  \tau_3, U \right] \; ,\nonumber \\
\chi&=&2\,B\,{\it M}\; ,
\end{eqnarray}
where ${\it M}$ denotes the SU(2) quark mass matrix in the isospin limit
$m_u=m_d$.

Finally, we emphasize that we do not require any
additional diagrams compared to the ${\cal O}(p^3)$ calculation for real
Compton scattering \cite{BKKM1}. The
complete set of non-zero diagrams we have to
calculate is given in Fig.\ref{figgborn} ((a) $s$-channel, (b)
$u$-channel, (c) contact diagram and (f) $t$-channel pole term)
and Fig.\ref{figgnpi} ($N\pi$-loop diagrams).
In the following we will treat the tree and loop parts of
the amplitudes separately,
\begin{equation}
A_i = A_i^{tree} + A_i^{loop} \, ,
\end{equation}
since the generalized polarizabilities are contained only in the latter.

\subsection{$\Delta$(1232) and the Small Scale Expansion}\label{deltatheory}

In standard SU(2) HBChPT, nucleon resonances like the $\Delta$(1232) are
considered to be much
heavier than the nucleon and therefore only contribute via local counterterms.
This
approach is particularly well-suited for near-threshold processes ({\it e.g.}
the
multipole $E_{0+}$ in threshold pion photoproduction) where the resonance
contributions
are small and their contribution to counterterms can be estimated by a simple
Born diagram analysis. However, if one wants to move away from threshold,
nucleon resonances, in particular the lowest lying SU(2) resonance
$\Delta$(1232),
contribute as dynamical degrees of freedom and the theoretical treatment in
terms
of local counterterms generates a slowly converging perturbative series. In
this kinematical
regime it is therefore advantageous to formulate an effective field theory
which keeps
the resonance as an explicit degree of freedom. In addition to this dynamical
consideration there is also another practical concern regarding the inclusion
of
resonance effects via counterterms. Even if simple Born exchange might be the
dominant
contribution of a particular resonance, the local counterterm in the chiral
Lagrangian
that subsumes this effect might be of higher order in the calculation, so
that the
leading and even the subleading result can misrepresent the perturbative
series. A
well-known example of this type are the so-called spin-polarizabilities of the
nucleon,
wherein one encounters very large contributions due to $\Delta$(1232) Born
graphs
that only start contributing via counterterms at ${\cal O}(p^5)$ in the
chiral
calculation ({\it e.g.} \cite{HHKK}). 
Situations of this type require a ``resummation'' of the
standard chiral expansion in order to push resonance effects into lower orders
to
restore meaningful perturbative expansions for quantities of interest in low
energy baryon physics.

In order to address these two different but related issues in the field of
resonance physics in
baryon CHPT, the so called ``small scale expansion'' of
SU(2) baryon ChPT has recently been formulated \cite{HBDel,HHK97}. In this
chiral
effective theory one treats the nucleon and
the first nucleon resonance---$\Delta$(1232)---as explicit degrees of freedom,
and, to address the second problem, the chiral power counting is modified to
bring $\Delta$(1232) related effects into lower orders of the calculation. In
the
``small scale expansion'' one organizes the Lagrangian and the calculation in
powers of the scale "$\epsilon$", which, in addition to the chiral expansion
parameters of small
momenta $q$ and the pion mass $m_\pi$, also includes the $\Delta(1232)-N(940)$ mass
splitting $\Delta=M_\Delta-M_N$.  Of course, this modification of the
chiral counting implies that one has to repeat the whole procedure of
construction of
the Lagrangian and the determination of counterterms and coupling
constants, even
for processes which only involve nucleons in the initial and final states.
For first
results regarding the modified renormalization of nucleon parameters we refer
to
\cite{bfhm}.

For our calculation below, which (as far as the GPs are concerned) 
is done only to leading
order---${\cal O}(\epsilon^3)$---in the small scale expansion of the
(generalized)
polarizabilities, we shall require only the propagator involving the
$\Delta(1232)$ as well as the couplings $NN\gamma$,
$NN\gamma\gamma$, $N\Delta\pi$ and $N\Delta\gamma$.  Details of the ``small
scale
expansion'' formalism are given in ref. \cite{HHK97}.  Here
we only list the minimal structures necessary for the present calculation.
The systematic 1/M-expansion of the coupled $N\Delta$-system starts with
the most general relativistic chiral invariant lagrangian
involving spin 1/2 ($\psi_N$) and spin 3/2 ($\psi^i_\mu$) baryon 
fields\footnote{In order to take into account the isospin 3/2
property of the $\Delta$(1232) we
supply the Rarita-Schwinger spinor with an additional isospin index $i$,
subject to the subsidiary condition $\tau_i \; \psi_{\mu}^i (x) = 0$.}.
The ``light'' spin 3/2 field $T^i_\mu$ in
the effective low-energy theory is projected from its relativistic
Rarita-Schwinger
counterpart $\psi^i_\mu$ via
\begin{equation}
T_{\mu}^i (x) \equiv  P_{v}^{+} \; P^{3/2}_{(33)\mu\nu} \; \psi^{\nu}_i (x)
                    \; \mbox{exp}(i M_0 v \cdot x), \label{eq:T}
\end{equation}
where we have introduced a spin 3/2 projection operator for fields with
{\it fixed velocity} $v_\mu$
\begin{equation}
P^{3/2}_{(33)\mu \nu} = g_{\mu \nu} - \frac{1}{3} \gamma_{\mu} \gamma_{
                        \nu} - \frac{1}{3} \left( \not\!{v} \gamma_{\mu}
                        v_{\nu} + v_{\mu} \gamma_{\nu}\not\!{v}\right) .
                        \label{eq:proj}
\end{equation}
The remaining components,
\begin{equation}
G_{\mu}^i (x)  =  \left( g_{\mu\nu}-P_{v}^{+} \; P^{3/2}_{(33)\mu\nu}\right)
                   \psi^{\nu}_i (x) \; \mbox{exp}(i M_0 v \cdot x) ,
\label{eq:G}
\end{equation}
can be shown to be ``heavy'' \cite{HHK97} and are integrated out. Resulting
from this
procedure one finds the (non-relativistic) chiral lagrangians of the ``small
scale
expansion'' (SSE):
\begin{equation}
{\cal L}^{SSE}={\cal L}_{N}^{SSE}+{\cal L}_{\Delta}^{SSE}+\left({\cal L}_{
N\Delta}^{SSE} + h.c. \right). \label{eq:4L}
\end{equation}
To the order we are working here ${\cal L}_{N}^{SSE}$ agrees with the chiral
lagrangian ${\cal L}_{N}^{(3)}$ (Eq.(\ref{eq:chiL})) needed for VCS. From
the
chiral SSE lagrangians explicitly involving the $\Delta$ field we need the
structures
\cite{HHK97}
\begin{eqnarray}
{\cal L}^{(1)}_{\Delta}&=&-\bar{T}^\mu_i g_{\mu\nu}\left[iv\cdot
D^{ij}-\Delta_0\;
                             \delta^{ij}+\ldots\right] T^\nu_j\nonumber\\
{\cal L}^{(1)}_{N\Delta}&=&g_{\pi N\Delta}\bar{T}^\mu_i\;w_\mu^i\;N+{\rm
                               h.c.}\nonumber\\
{\cal L}^{(2)}_{N\Delta}&=&\bar{T}^\mu_i\left[{ib_1\over
M_0}\;S^\nu\;f_{+\mu\nu}^i+
                           \ldots \right] N +{\rm h.c.}, \label{eq:xxx}
\end{eqnarray}
where $\Delta_0=M_\Delta-M_0$ can be identified with the {\it physical}
delta-nucleon
mass difference to the order we are working, {\it i.e.} $M_0\equiv M_N$.
The corresponding chiral tensors needed for VCS read
\begin{eqnarray}
D_{\mu}^{ij}&=&\partial_{\mu}\delta^{ij}-i\frac{e}{2}\left(1+\tau_3\right)A_\mu
               \delta^{ij}+e\epsilon^{i3j}A_\mu + \ldots \nonumber\\
w_\mu^i
    &=&-\frac{1}{F_\pi}\partial_\mu\pi^i-\frac{e}{F_\pi}A_\mu\epsilon^{i3j}
               \pi^j+\ldots \nonumber\\
f_{+\mu\nu}^i&=&e\delta^{i3}\left(\partial_\mu A_\nu-\partial_\nu
A_\mu\right)+\dots
\end{eqnarray}
The coupling constants defined in Eq.(\ref{eq:xxx}) are determined from fits to
the strong and electromagnetic decay widths of the Delta resonance within the
``small
scale
expansion''. To the order we are working one requires\footnote{Note
that these values are determined from the width expressions within the ``small
scale
expansion" and therefore differ from those obtained in a relativistic analysis,
{\it e.g.} see ref. \cite{delta}.} \cite{HHKK,trh}
$g_{\pi N\Delta}=1.05\pm 0.02$ and $b_1=3.85\pm 0.15$.

The leading propagator for a $\Delta$(1232) field with small momentum $k_\mu$ is 
then given by
\begin{equation}
S^{3/2}_{\mu\nu}={-iP^{3/2}_{\mu\nu}\over v\cdot
k-\Delta+i\eta}\xi^{ij}_{I=3/2},
\end{equation}
where $P^{3/2}_{\mu\nu}$ is the spin-${3\over 2}$ heavy baryon projector in
d-dimensions \cite{HHK97}
\begin{equation}
P^{3/2}_{\mu\nu}=g_{\mu\nu}-v_\mu v_\nu+{4\over d-1}S_\mu S_\nu ,
\end{equation}
and
\begin{equation}
\xi^{ij}_{I=3/2}=\delta^{ij}-{1\over 3}\tau^i\tau^j
\end{equation}
is the corresponding isospin projector.
The vertices relevant for our calculation can be read off directly from
Eq.(\ref{eq:xxx}).  As in the nucleon case, the resulting diagrams
can be separated into two classes---one-loop graphs and Born graphs. The
systematics
of the ``small scale expansion'' uniquely fixes the number and type of diagrams
for
VCS to be calculated to ${\cal O}(\epsilon^3)$. It turns out that to the order
we
are working there are two Born diagrams involving the $\Delta$(1232)
(Fig.\ref{figgborn}(d,e)) and nine $\Delta\pi$-loop diagrams
(Fig.\ref{figgdpi}),
which turn out to have exactly the same
structure as
their chiral $N\pi$ analogues (cf. Fig.\ref{figgnpi}). However, before
undertaking
any such calculation, it is necessary to work out the formalism for VCS.

\section{Virtual Compton Scattering}
\subsection{General Structure} \label{svcs}
We begin by specifying our notation for the virtual Compton process
\begin{equation}
\gamma^*(\epsilon^\mu,q^\mu)+N(p_i^\mu)\rightarrow\gamma(\epsilon^{\prime
*\mu},q^{
\prime\mu})+N(p_f^\mu).
\end{equation}
Here the nucleon four-momenta in the initial and final states are denoted by
$p_i^\mu=(E_i,\vec{p}_i)$ and $p_f^\mu=(E_f,\vec{p}_f)$ respectively. The
virtual initial [real final] state photon is characterized by its four-momentum
$q^\mu=(\omega,\vec{q}),\; q^2<0$
[$q^{\prime\mu}=(\omega^\prime,\vec{q}^{\;\prime}),\;q^{\prime 2}=0$] and
polarization vector $\epsilon^\mu=(\epsilon_0,\vec{\epsilon})$
[$\epsilon^{\prime\mu}=(\epsilon^{\prime}_0,\vec{\epsilon}^{\;\prime})$].

Since our discussion refers to an electron scattering experiment, wherein the
virtual photon is exchanged between the electron and hadron currents, the
polarization vector of the incoming photon is given by
\begin{equation}
\epsilon_\mu=e\;\bar{u}_{e'}(k_1)\;\gamma_\mu \;u_e(k_2)\; \frac{1}{q^2} \; ,
\end{equation}
where $u_e(k_1),\bar{u}_{e'}(k_2)$ are electron Dirac spinors with four-momenta
$k_1^\mu \;(k_2^{\mu})$ before (after) emission of the virtual photon.
The unit charge $e$ is taken as $e=\sqrt{4\pi/137}>0$.

In addition to the proper VCS process displayed in Fig.2a there are also
Bethe-Heitler processes taking place (Fig.2b,c),
\begin{equation}
{\cal M}_{eN\rightarrow e'N\gamma}={\cal M}^{VCS}+{\cal M}^{Bethe-Heitler},
\end{equation}
and such Bethe-Heitler contributions must be carefully evaluated before one can
infer any
information about the VCS matrix element from the electron scattering cross
section.\footnote{In fact, the primary source of information about the
structure of the
nucleon in the process $eN\rightarrow e'N\gamma$ comes from the interference
between ${\cal M}^{VCS}$ and ${\cal M}^{Bethe-Heitler}$.}
In the following, however, we will focus on the evaluation of the VCS matrix
element
${\cal M}^{VCS}$ (Fig.2a). For details on ${\cal M}^{Bethe-Heitler}$ and the
calculation
of the cross section we refer to \cite{nstar} and references
therein.

{From} now on we will work in the center of mass system of the final state
photon-nucleon subsystem,
\begin{eqnarray}
\vec{p}_f=-\vec{q}^{\;\prime},& &
\vec{p}_i=-\vec{q}=-\bar{q}\;\hat{e}_z\; , \nonumber \\
\omega^\prime+\sqrt{M_{N}^2+\omega^{\prime 2}}&=&\omega
+\sqrt{M_{N}^2+\bar{q}^2} \; ,
\label{eq:omega}
\end{eqnarray}
where the $z$-axis is defined by the three-momentum vector $\vec{q}$ of the 
incoming virtual photon.  Utilizing the Lorentz gauge\footnote{Our
calculations are actually performed in the Coulomb gauge, see the discussion in appendix
\ref{appnpi}.},
\begin{equation}
\epsilon \cdot q=0,\qquad \epsilon_0=\frac{\bar{q}}{\omega}\;
                                                      \epsilon_z \; ,
\end{equation}
with $\vec{\epsilon}=\vec{\epsilon}_T+\epsilon_z \hat{e}_z$,
one can express the VCS matrix element in terms of
twelve\footnote{It is helpful to employ the identity given in appendix \ref{ident}
when reducing Pauli structures to the 12 structure amplitudes employed here.}
independent kinematic forms
\begin{eqnarray}
{\cal M}^{VCS}&=&i\;e^2\left\{\vec{\epsilon}^{\;\prime
*}\cdot\vec{\epsilon}_T\;A_1
+\vec{\epsilon}^{\;\prime
*}\cdot\hat{q}\;\vec{\epsilon}_T\cdot\hat{q}^\prime\;A_2+i
\vec{\sigma}\cdot\left(\vec{\epsilon}^{\;\prime
*}\times\vec{\epsilon}_T\right)A_3+i\;\vec{
\sigma}\cdot\left(\hat{q}^\prime\times\hat{q}\right)\vec{\epsilon}^{\;\prime
*}\cdot
\vec{\epsilon}_T\;A_4 \right. \nonumber \\
& &\phantom{i\;e^2 }+i\;\vec{\sigma}\cdot\left(\vec{\epsilon}^{\;\prime
*}\times\hat{q}
\right)\vec{\epsilon}_T\cdot\hat{q}^\prime\;A_5+i\;\vec{\sigma}\cdot\left(\vec{
\epsilon}^{\;\prime
*}\times\hat{q}^\prime\right)\vec{\epsilon}_T\cdot\hat{q}^\prime\;
A_6 \nonumber \\
& &\phantom{i\;e^2}-i\;\vec{\sigma}\cdot\left(\vec{\epsilon}_T\times\hat{q}^\prime
\right)\vec{\epsilon}^{\;\prime
*}\cdot\hat{q}\;A_7-i\;\vec{\sigma}\cdot\left(\vec{
\epsilon}_T\times\hat{q}\right)\vec{\epsilon}^{\;\prime *}\cdot\hat{q}\;A_8
\nonumber \\
& &\phantom{i\;e^2
}\left.+\frac{q^2}{\omega^2}\;\epsilon_z\left[\vec{\epsilon}^{\;
\prime
*}\cdot\hat{q}\;A_9+i\;\vec{\sigma}\cdot\left(\hat{q}^\prime\times\hat{q}
\right)\vec{\epsilon}^{\;\prime
*}\cdot\hat{q}\;A_{10}+i\;\vec{\sigma}\cdot\left(
\vec{\epsilon}^{\;\prime
*}\times\hat{q}\right)A_{11}+i\;\vec{\sigma}\cdot\left(\vec{
\epsilon}^{\;\prime *}\times\hat{q}^\prime\right)A_{12}\right]\right\},
\label{eq:vcs12}
\end{eqnarray}
where $\sigma_i,\;i=x,y,z$ are Pauli spin matrices. Utilizing
Eq.(\ref{eq:omega}), each
amplitude $A_i$, i=1,12 is then a function of three independent
kinematic quantities---$\omega',\bar{q}$ and $\theta$.

\subsection{Separation of Born- and Structure-Part}\label{separation}
The twelve VCS amplitudes $A_i(\omega^\prime,\theta,\bar{q})$ can be decomposed
into a
(nucleon) Born part $A^{Born}_i(\omega^\prime,\theta,\bar{q})$ and a structure
dependent part $\bar{A}_i(\omega^\prime,\theta,\bar{q})$,
\begin{equation}
A_i(\omega^\prime,\theta,\bar{q})=A^{Born}_i(\omega^\prime,\theta,\bar{q})+
                                  \bar{A}_i(\omega^\prime,\theta,\bar{q}).
\end{equation}
To third order in both the chiral and small scale
expansions, the Born part contains the nucleon pole diagrams 
(Fig.\ref{figgborn}(a,b)), the Thomson seagull
graph (Fig.\ref{figgborn}(c)) and the (anomalous) pion-pole graph
(Fig.\ref{figgborn}(f)). In the case of a
proton target one finds
\begin{eqnarray}
A_1^{Born\;(3)}(\omega^\prime,\theta,\bar{q})&=&- \frac{1}{M_N}+{\cal
O}(1/(M_{N}^3,
               \Lambda_{\chi}^2M_N)) \nonumber\\
A_2^{Born\;(3)}(\omega^\prime,\theta,\bar{q})&=&\frac{\bar{q}}{M_{N}^2}+{\cal
O}(
               1/(M_{N}^3,\Lambda_{\chi}^2M_N)) \nonumber\\
A_3^{Born\;(3)}(\omega^\prime,\theta,\bar{q})&=&\frac{\left(1+2\kappa_p\right)
               \omega^\prime-\left(1+\kappa_p\right)^2\cos\theta\;\bar{q}}{2M_{N}^2}
               -{g_A\over
8\pi^2F_\pi^2}\;\frac{\omega^\prime\left(\omega^{\prime 2}+
               \bar{q}^2-2\omega^{\prime}\bar{q}\cos\theta\right)}{m_\pi^2+\omega^{
               \prime 2}+\bar{q}^2-2\omega^\prime\bar{q}\cos\theta} \nonumber\\
            & &\phantom{\frac{\left(1+2\kappa_p\right)
               \omega^\prime-\left(1+\kappa_p\right)^2\cos\theta\;\bar{q}}{2M_{N}^2}}
               +{\cal O}(1/(M_{N}^3,\Lambda_{\chi}^2M_N)) \nonumber\\
A_4^{Born\;(3)}(\omega^\prime,\theta,\bar{q})&=&-\frac{\bar{q}\left(1+\kappa_p\right)^2}
               {2 M_{N}^2}+{\cal O}(1/(M_{N}^3,\Lambda_{\chi}^2M_N))
\nonumber\\
A_5^{Born\;(3)}(\omega^\prime,\theta,\bar{q})&=&\frac{\bar{q}\left(1+\kappa_p\right)^2}
               {2 M_{N}^2}-{g_A\over 8\pi^2F_\pi^2}\;\frac{\omega^{\prime
2}\bar{q}}
               {m_\pi^2+\omega^{\prime
2}+\bar{q}^2-2\omega^\prime\bar{q}\cos\theta}+
               {\cal O}(1/(M_{N}^3,\Lambda_{\chi}^2M_N)) \nonumber\\
A_6^{Born\;(3)}(\omega^\prime,\theta,\bar{q})&=&-\frac{\omega^\prime\left(1+\kappa_p
               \right)}{2 M_{N}^2}+{g_A\over
8\pi^2F_\pi^2}\;\frac{\omega^{\prime 3}}
               {m_\pi^2+\omega^{\prime
2}+\bar{q}^2-2\omega^\prime\bar{q}\cos\theta}+
               {\cal O}(1/(M_{N}^3,\Lambda_{\chi}^2M_N)) \nonumber\\
A_7^{Born\;(3)}(\omega^\prime,\theta,\bar{q})&=&\frac{\bar{q}\left(1+\kappa_p
               \right)^2}{2 M_{N}^2}-{g_A\over
8\pi^2F_\pi^2}\;\frac{\omega^{\prime 2}
               \bar{q}}{m_\pi^2+\omega^{\prime
2}+\bar{q}^2-2\omega^\prime\bar{q}\cos
               \theta}+{\cal O}(1/(M_{N}^3,\Lambda_{\chi}^2M_N)) \nonumber\\
A_8^{Born\;(3)}(\omega^\prime,\theta,\bar{q})&=&-\frac{1+\kappa_p}{2
M_{N}^2}\;\frac{
               \bar{q}^2}{\omega^\prime}+{g_A\over
8\pi^2F_\pi^2}\;\frac{\omega^\prime
               \bar{q}^2}{m_\pi^2+\omega^{\prime
2}+\bar{q}^2-2\omega^\prime\bar{q}\cos
               \theta}+{\cal O}(1/(M_{N}^3,\Lambda_{\chi}^2M_N)) \nonumber\\
A_9^{Born\;(3)}(\omega^\prime,\theta,\bar{q})&=&-\frac{1}{M_N}+\frac{2\;\omega^\prime
               \bar{q}\cos\theta+\bar{q}^2}{2 M_{N}^2\omega^\prime}+{\cal
O}(1/(M_{N}^3,
               \Lambda_{\chi}^2M_N)) \nonumber\\
A_{10}^{Born\;(3)}(\omega^\prime,\theta,\bar{q})&=&-{g_A\over
8\pi^2F_\pi^2}\;\frac{
                  \omega^{\prime 2}\bar{q}}{m_\pi^2+\omega^{\prime
2}+\bar{q}^2-2
                  \omega^\prime\bar{q}\cos\theta}+{\cal
O}(1/(M_{N}^3,\Lambda_{\chi}^2
                  M_N)) \nonumber\\
A_{11}^{Born\;(3)}(\omega^\prime,\theta,\bar{q})&=&\frac{\left(1+2\kappa_p\right)
                  \omega^\prime}{2 M_{N}^2}-{g_A\over
8\pi^2F_\pi^2}\;\frac{\omega^{
                  \prime
2}\left(\omega^\prime-\bar{q}\cos\theta\right)}{m_\pi^2+
                  \omega^{\prime
2}+\bar{q}^2-2\omega^\prime\bar{q}\cos\theta}+{\cal O}
                  (1/(M_{N}^3,\Lambda_{\chi}^2M_N))\nonumber\\
A_{12}^{Born\;(3)}(\omega^\prime,\theta,\bar{q})&=&-\frac{\left(1+\kappa_p\right)
                  \omega^\prime\cos\theta}{2 M_{N}^2}-{g_A\over
8\pi^2F_\pi^2}\;\frac{
                  \omega^{\prime
2}\left(\bar{q}-\omega^\prime\cos\theta\right)}{m_\pi^2
                  +\omega^{\prime
2}+\bar{q}^2-2\omega^\prime\bar{q}\cos\theta}+{\cal O}
                  (1/(M_{N}^3,\Lambda_{\chi}^2M_N)) \,,\label{eq:bg}
\end{eqnarray}
where $\Lambda_\chi=4\pi F_\pi$ denotes the scale of chiral symmetry
breaking\cite{sca}.  One can easily verify that the low energy forms of these
structure
functions are in agreement with the constraints implied by the Low theorem in
the case of real Compton scattering\cite{low}---$\bar{q}=0$---and with the
generalized low energy theorem in the case of VCS\cite{Guichon,SK}.
{From} the above expressions it can also
be seen that the pion-pole contributions---Fig.\ref{figgborn}(f)---which
scale linearly with $g_A$, affect only
the spin-dependent structure amplitudes, as expected from the
pion-nucleon coupling structure.
All additional contributions are contained in the structure-dependent parts
$\bar{A}_i(\omega^\prime,\theta,\bar{q})$ of the amplitudes, from which one can
extract the (generalized) polarizabilities.

\subsection{Connection with the GPs}
In this section we present the formulae by which the GPs are related to the
twelve
structure-dependent amplitudes
$\bar{A}_i(\omega^\prime,\theta,\bar{q}),\;i=1\dots 12$
{\em to ${\cal O}(p^3)$ in HBChPT and to ${\cal O}(\epsilon^3)$ in
SSE}.  First, we focus on the spin-independent GPs.

To leading order in both the chiral and small scale expansions the
spin-independent GPs $\bar{\alpha}_E(\bar{q}), \; \bar{\beta}_M(\bar{q})$ can
be found from the structure functions
$\bar{A}_9(\omega^\prime,\theta,\bar{q}),\;\bar{A}_2(\omega^\prime,\theta,\bar{q})$
via \cite{HHKS1}
\begin{eqnarray}
\bar{\alpha}^{(3)}_E(\bar{q})&=&\frac{e^2}{8\pi} \; \frac{\partial^2}{\partial
\omega^{\prime 2}}\bar{A}^{(3)}_9(\omega^\prime,\theta,\bar{q})\biggl|_{\omega^\prime
=0}
\; , \nonumber\\
\bar{\beta}^{(3)}_M(\bar{q})&=&-\frac{e^2}{4\pi}
\;\frac{1}{\bar{q}}\;\frac{\partial}{
\partial\omega^\prime}\bar{A}^{(3)}_2(\omega^\prime,\theta,\bar{q})\biggl|_{\omega^\prime
=0} \; . \label{defab}
\end{eqnarray}
Note that the structure amplitudes in general have a dependence on the
scattering angle
$\theta$, whereas the GPs are only functions of $\bar{q}$. The independence of
the
GPs on $\theta$ therefore serves as a non-trivial check on the calculation.

Likewise, the four independent spin-dependent GPs can be found from the
relations
\cite{Germar}
\begin{eqnarray}
\hat{P}_{(01,1)1}^{(3)}(\bar{q})&=&-\frac{\sqrt{2}}{3\sqrt{3}}\;\frac{1}{\bar{q}^2}\;
\frac{\partial}{\partial\omega^\prime}\left[2\;\bar{A}_{3}^{(3)}(\omega^\prime,\theta,
\bar{q})+\bar{A}_{8}^{(3)}(\omega^\prime,\theta,\bar{q})\right]_{\omega^\prime=0}
\nonumber\\
P_{(01,12)1}^{(3)}(\bar{q})&=&-\frac{\sqrt{2}}{3}\;\frac{1}{\bar{q}^2}\;\frac{\partial}{
\partial\omega^\prime}\bar{A}_{8}^{(3)}(\omega^\prime,\theta,\bar{q})\biggl|_{\omega^\prime=0}
\nonumber\\
P_{(11,02)1}^{(3)}(\bar{q})&=&-\frac{\sqrt{2}}{3\sqrt{3}}\;\frac{1}{\bar{q}}\;\frac{
\partial^2}{\partial\omega^{\prime2}}\bar{A}_{10}^{(3)}(\omega^\prime,\theta,\bar{q})\biggl|_{
\omega^\prime=0}
\nonumber\\
P_{(11,00)1}^{(3)}(\bar{q})&=&\frac{\bar{q}}{\sqrt{3}}\;\frac{\partial^2}{\partial
\omega^{\prime
2}}\left[\bar{A}_{12}^{(3)}(\omega^\prime,\theta,\bar{q})-\frac{2}{3}\;
\bar{A}_{10}^{(3)}(\omega^\prime,\theta,\bar{q})\right]_{\omega^\prime=0}.\label{defspin}
\end{eqnarray}
We note that these relations are only exact to third order in the chiral and in
the small scale expansion. The analysis of ref.\cite{Germar} must be
generalized
before one can perform any fourth order calculations. Thus, to the order we are
working,
the remaining three spin-dependent GPs
$P_{(01,01)1}^{(3)},P_{(11,11)1}^{(3)},\hat{P}_{(11,2)1}^{(3)}$ and the
additional scalar GP $\hat{P}_{(01,1)0}^{(3)}$ can only be
reconstructed\footnote{The
origin of this impediment lies in the fact that the quantity
$\omega_0=M_N-\sqrt{M_{N}^2+\bar{q}^2}$
strictly speaking is suppressed by a factor of $1/M_N$ in both the chiral and
small scale expansions. Full sensitivity to $\omega_0$ dependent
quantities can
therefore only be achieved in ${\cal O}(p^4)$, respectively ${\cal
O}(\epsilon^4)$
calculations.
} with the help of the charge-conjugation constraint of
Eqs.(\ref{cinv}), yielding
\begin{eqnarray}
\hat{P}_{(01,1)0}^{(3)}(\bar{q})&=&\frac{\omega_0}{3\;\bar{q}^2}\left[\frac{\partial^2}{
\partial\omega^{\prime
2}}\bar{A}^{(3)}_9(\omega^\prime,\theta,\bar{q})-\frac{2}{
\bar{q}}\;\frac{\partial}{\partial\omega^\prime}\bar{A}^{(3)}_2(\omega^\prime,\theta,
\bar{q})\right]_{\omega^\prime =0}
\nonumber\\
P_{(01,01)1}^{(3)}(\bar{q})&=&\frac{\omega_0}{3\;\bar{q}}\;\frac{\partial^2}{\partial
\omega^{\prime
2}}\left[\bar{A}_{12}^{(3)}(\omega^\prime,\theta,\bar{q})-\bar{A}_{10}^{
(3)}(\omega^\prime,\theta,\bar{q})\right]_{\omega^\prime =0}
\nonumber\\
P_{(11,11)1}^{(3)}(\bar{q})&=&\frac{\omega_0}{3\;\bar{q}^2}\left\{2\;\frac{\partial}{
\partial\omega^\prime}\bar{A}_{3}^{(3)}(\omega^\prime,\theta,\bar{q})\biggl|_{\omega^\prime=0}
-\frac{\omega_{0}^2}{\bar{q}}\;\frac{\partial^2}{\partial\omega^{\prime
2}}\left[
\bar{A}_{12}^{(3)}(\omega^\prime,\theta,\bar{q})-\bar{A}_{10}^{(3)}(\omega^\prime,
\theta,\bar{q})\right]_{\omega^\prime=0}\right\}
\nonumber\\
\hat{P}_{(11,2)1}^{(3)}(\bar{q})&=&\frac{\sqrt{2}\;\omega_0}{3\sqrt{5}\;\bar{q}^3}
\left\{\left[\frac{\omega_{0}^2}{\bar{q}^2}\;\frac{\partial^2}{\partial\omega^{
\prime
2}}\bar{A}_{12}^{(3)}(\omega^\prime,\theta,\bar{q})+\left(1-\frac{\omega_{0}^2}{
\bar{q}^2}\right)\frac{\partial^2}{\partial\omega^{\prime
2}}\bar{A}_{10}^{(3)}(
\omega^\prime,\theta,\bar{q})\right]_{\omega^\prime=0}\right. \nonumber \\
& &\phantom{\frac{\sqrt{2}\;\omega_0}{3\sqrt{5}\;\bar{q}^3} }\left.
   -\frac{2}{\bar{q}}\frac{\partial}{\partial\omega^\prime}\bar{A}_{3}^{(3)}(
   \omega^\prime,\theta,\bar{q})\biggl|_{\omega^\prime=0}\right\} , \label{defcharge}
\end{eqnarray}
with $\omega_0=M_N-\sqrt{M_{N}^2+\bar{q}^2}$. Note that the spin-dependent GPs
are just functions of the three-momentum transfer $\bar{q}$, whereas their
generating
structure amplitudes in Eqs.(\ref{defspin}-\ref{defcharge}) also depend on the
scattering angle $\theta$---leading again to a non-trivial check on the
calculation
as in the case of the spin-independent GPs.

With these definitions of the GPs we now turn to the results of
the chiral and small scale expansions.

\section{Results}
In this section we present the results for the generalized polarizabilities
calculated in {\em two different chiral effective theories}---${\cal O}(p^3)$
HBChPT
and ${\cal O}(\epsilon^3)$ SSE.
\subsection{${\cal O}(p^3)$ Heavy Baryon ChPT}
\subsubsection{Structure Amplitudes} \label{chiamp}

The only diagrams left at ${\cal O}(p^3)$ for the structure dependent part are
the
nine $N\pi$-continuum diagrams (Fig.\ref{figgnpi}), which correspond to the
pion-cloud of the nucleon in
the formalism of baryon chiral perturbation theory. All other diagrams have
already been
accounted for in the Born part of section \ref{separation}. We can now
calculate
the ${\cal O}(p^3)$ contributions to the 12
VCS structure amplitudes defined in Eq.(\ref{eq:vcs12}), with all our results
given in the CMS
of the the final state photon-nucleon subsystem. From appendix \ref{appnpi} 
one can read off the
spin-independent structure amplitudes to ${\cal O}(p^3)$, yielding
\begin{eqnarray}
\bar{A}^{(3)}_1(\omega^\prime,\theta,\bar{q})&=&
-\frac{g_{A}^2}{16\pi
F_{\pi}^2}\int_{0}^{1}dx\int_{0}^{1}dy\left\{\frac{m_{\pi}^2-4
m_{f}^2}{\sqrt{m_{f}^2}}+2\sqrt{m_{\pi}^2-\omega^{\prime
2}}-2\sqrt{m_{\pi}^2-\omega^{
\prime 2}x^2}-2\sqrt{\tilde{m}^2-\omega^{\prime 2}x^2} \right. \nonumber \\
& & \phantom{-\frac{g_{A}^2}{16\pi F_{\pi}^2}\int_{0}^{1}dx\int_{0}^{1}dy
}\left.+2\left(1-y\right)\frac{K^2 T+\left(6\hat{m}^2-m_{\pi}^2-6
T^2\right)\omega^\prime}{\omega^\prime\sqrt{\hat{m}^2-T^2}}\right\}
\nonumber\\
\bar{A}^{(3)}_2(\omega^\prime,\theta,\bar{q})&=&
+\frac{g_{A}^2}{8\pi F_{\pi}^2}\int_{0}^{1}dx\int_{0}^{1}dy \frac{\bar{q}
\omega^\prime\left(1-y\right) }{\sqrt{\hat{m}^2-T^2}}\left\{-1+x-8x
y+7\left(y-y^2+x y^2\right) \right. \nonumber \\
& & \phantom{-\frac{g_{A}^2}{8\pi F_{\pi}^2}\int_{0}^{1}dx\int_{0}^{1}dy
\frac{\bar{q} \omega^\prime\left(1-y\right) }{\sqrt{\hat{m}^2-T^2}} } \left.
+\left(1-x\right)y\left(1-y\right)\frac{\left(m_{\pi}^2-\hat{m}^2+T^2\right)
\omega^\prime-K^2 T}{\omega^\prime\left(\hat{m}^2-T^2\right)}\right\}
\nonumber\\
\bar{A}^{(3)}_9(\omega^\prime,\theta,\bar{q})&=&
\bar{A}^{(3)}_1(\omega^\prime,\theta,\bar{q})+\cos\theta\;\bar{A}^{(3)}_2(
\omega^\prime,\theta,\bar{q})+ \frac{g_{A}^2}{16\pi
F_{\pi}^2}\int_{0}^{1}dx\int_{0}^{
1}dy\;\bar{q}^2\left[\frac{x\left(1-2x\right)}{\sqrt{\tilde{m}^2-\omega^{\prime
2}}} \right. \nonumber \\
& &\left.
-\left(1-2y\right)y\left(1-y\right)\frac{\left(m_{\pi}^2-\hat{m}^2+T^2\right)
\omega^\prime-K^2 T}{\omega^\prime\left(\hat{m}^2-T^2\right)^{3/2}}+
\frac{\left(1-y\right)\left(1-9y+14y^2\right)}{\sqrt{\hat{m}^2-T^2}}\right] ,
\label{eq:a1a2}
\end{eqnarray}
with the ``energy'' and ``mass'' variables
\begin{eqnarray}\label{definitions}
T&=&\omega^\prime x\left(1-y\right) \nonumber\\
K^2&=&\omega^{\prime 2}-\omega^\prime\bar{q}\cos\theta \nonumber\\
\tilde{m}^2&=&m_{\pi}^2-q^2 x\left(1-x\right) \nonumber\\
\hat{m}^2&=&m_{\pi}^2-q^2 y\left(1-y\right)+2 q\cdot q^\prime\left(1-x\right)y
            \left(1-y\right) \nonumber\\
m_{f}^2&=&m_{\pi}^2-\left(q-q^\prime\right)^2 x\left(1-x\right) .
\end{eqnarray}
The spin-dependent structure amplitudes to ${\cal O}(p^3)$ in the chiral
expansion can
also be found from the expressions in appendix \ref{appnpi}---
\begin{eqnarray}
\bar{A}^{(3)}_3(\omega^\prime,\theta,\bar{q})&=& \frac{g_{A}^2}{4\pi^2
F_{\pi}^2}\int_{0}^{1}dx\int_{0}^{1}dy\left\{-\sqrt{m_{\pi}^2-\omega^{\prime
2}}\arcsin\left[\frac{\omega^\prime}{m_\pi}\right]+\sqrt{m_{\pi}^2-\omega^{\prime
2}x^2}\arcsin\left[\frac{\omega^\prime x}{m_\pi}\right] \right. \nonumber \\
& &\phantom{ \frac{g_{A}^2}{4\pi^2 F_{\pi}^2}\int_{0}^{1}dx\int_{0}^{1}dy }
+\sqrt{\tilde{m}^2-\omega^{\prime 2} x^2}\arcsin\left[\frac{\omega^\prime
x}{\tilde{m}}\right] +\omega^\prime x\log\left[\frac{\tilde{m}}{m_{\pi}}\right]
\nonumber\\
& &\phantom{ \frac{g_{A}^2}{4\pi^2 F_{\pi}^2}\int_{0}^{1}dx\int_{0}^{1}dy }
\left.
+\sin^2\theta\left(1-x\right)x\left(1-y\right)^3y\;\frac{\bar{q}^2\omega^{\prime
2}\left(T\sqrt{\hat{m}^2-T^2}+\hat{m}^2\arcsin\left[\frac{T}{\hat{m}}\right]\right)}{
\hat{m}^2\left(\hat{m}^2-T^2\right)^{3/2}} \right\}
\nonumber\\
\bar{A}^{(3)}_4(\omega^\prime,\theta,\bar{q})&=& \frac{g_{A}^2}{4\pi^2
F_{\pi}^2}\int_{0}^{1}dx\int_{0}^{1}dy\left(1-y\right)\frac{\bar{q}\;
T}{\sqrt{\hat{m}^2-T^2}}\arcsin\left[\frac{T}{\hat{m}}\right]
\nonumber\\
\bar{A}^{(3)}_5(\omega^\prime,\theta,\bar{q})&=& \frac{g_{A}^2}{4\pi^2
F_{\pi}^2}\int_{0}^{1}dx\int_{0}^{1}dy\left\{\left(x-1\right)\left(1-y\right)^2\frac{
\bar{q}\;\omega^\prime\arcsin\left[\frac{T}{\hat{m}}\right]}{\sqrt{\hat{m}^2-T^2}}
\right. \nonumber \\
& &\phantom{ \frac{g_{A}^2}{4\pi^2
F_{\pi}^2}\int_{0}^{1}dx\int_{0}^{1}dy}\left.
+\cos\theta\left(1-x\right)x\left(1-y\right)^3y\;\frac{\bar{q}^2\omega^{\prime
2}\left(T\sqrt{\hat{m}^2-T^2}+\hat{m}^2\arcsin\left[\frac{T}{\hat{m}}\right]\right)}{
\hat{m}^2\left(\hat{m}^2-T^2\right)^{3/2}} \right\}
\nonumber\\
\bar{A}^{(3)}_6(\omega^\prime,\theta,\bar{q})&=& \frac{g_{A}^2}{4\pi^2
F_{\pi}^2}\int_{0}^{1}dx\int_{0}^{1}dy\left\{\left(1-x\right)\left(1-y\right)^2\frac{
\omega^{\prime 2}\arcsin\left[\frac{T}{\hat{m}}\right]}{\sqrt{\hat{m}^2-T^2}}
\right.
\nonumber \\
& &\phantom{ \frac{g_{A}^2}{4\pi^2
F_{\pi}^2}\int_{0}^{1}dx\int_{0}^{1}dy}\left.
-\left(1-x\right)x\left(1-y\right)^3y\;\frac{\bar{q}^2\omega^{\prime
2}\left(T\sqrt{\hat{m}^2-T^2}+\hat{m}^2\arcsin\left[\frac{T}{\hat{m}}\right]\right)}{
\hat{m}^2\left(\hat{m}^2-T^2\right)^{3/2}} \right\}
\nonumber\\
\bar{A}^{(3)}_7(\omega^\prime,\theta,\bar{q})&=&  \frac{g_{A}^2}{4\pi^2
F_{\pi}^2}\int_{0}^{1}dx\int_{0}^{1}dy\left\{y\left(y-1\right)\frac{\omega^\prime
\bar{q}\arcsin\left[\frac{T}{\hat{m}}\right]}{\sqrt{\hat{m}^2-T^2}}
\right. \nonumber \\
& &\phantom{ \frac{g_{A}^2}{4\pi^2
F_{\pi}^2}\int_{0}^{1}dx\int_{0}^{1}dy}\left.
+\cos\theta\left(1-x\right)x\left(1-y\right)^3y\;\frac{\bar{q}^2\omega^{\prime
2}\left(T\sqrt{\hat{m}^2-T^2}+\hat{m}^2\arcsin\left[\frac{T}{\hat{m}}\right]\right)}{
\hat{m}^2\left(\hat{m}^2-T^2\right)^{3/2}} \right\}
\nonumber\\
\bar{A}^{(3)}_8(\omega^\prime,\theta,\bar{q})&=&  \frac{g_{A}^2}{4\pi^2
F_{\pi}^2}\int_{0}^{1}dx\int_{0}^{1}dy\left\{y\left(1-y\right)\frac{\bar{q}^{2}
\arcsin\left[\frac{T}{\hat{m}}\right]}{\sqrt{\hat{m}^2-T^2}}
\right. \nonumber \\
& &\phantom{ \frac{g_{A}^2}{4\pi^2
F_{\pi}^2}\int_{0}^{1}dx\int_{0}^{1}dy}\left.
-\left(1-x\right)x\left(1-y\right)^3y\;\frac{\bar{q}^2\omega^{\prime
2}\left(T\sqrt{\hat{m}^2-T^2}+\hat{m}^2\arcsin\left[\frac{T}{\hat{m}}\right]\right)}{
\hat{m}^2\left(\hat{m}^2-T^2\right)^{3/2}} \right\}
\nonumber\\
\bar{A}^{(3)}_{10}(\omega^\prime,\theta,\bar{q})&=&\bar{A}^{(3)}_{4}(\omega^\prime,
\theta,\bar{q})+\bar{A}^{(3)}_{7}(\omega^\prime,\theta,\bar{q})\nonumber \\
& &\phantom{\bar{A}^{(3)}_{4}(\omega^\prime,\theta,\bar{q})}
+\frac{g_{A}^2}{8\pi^2
F_{\pi}^2}\int_{0}^{1}dx\int_{0}^{1}dy\left(1-y\right)^2
x\left(2y-1\right)y\;\frac{
\bar{q}^3\omega^{\prime}\left(T\sqrt{\hat{m}^2-T^2}+\hat{m}^2\arcsin\left[\frac{T}{
\hat{m}}\right]\right)}{\hat{m}^2\left(\hat{m}^2-T^2\right)^{3/2}}
\nonumber\\
\bar{A}^{(3)}_{11}(\omega^\prime,\theta,\bar{q})&=& A_3(\omega',\theta,\bar{q})
+A_5(\omega',\theta,\bar{q})\nonumber\\
&+&{g_A^2\over 8\pi^2F_\pi^2}\int_0^1 dx \int_0^1 dy\left\{
x(1-2x)\bar{q}^2{\arcsin[{\omega'x\over \tilde{m}}]\over
\sqrt{\tilde{m}^2-{\omega'}^2x^2}}+(1-y)(1-2y)\bar{q}^2
{\arcsin[{T\over \hat{m}}]\over \sqrt{\hat{m}^2-T^2}}\right.\nonumber\\
& &\phantom{sin}\left.-x(1-x)y(1-y)^3\bar{q}^3\omega'{T\sqrt{\hat{m}^2-T^2}
+\hat{m}^2\arcsin[{T\over \hat{m}}]\over \hat{m}^2(\hat{m}^2-T^2)^{3\over
2}}\right\}
\nonumber\\
\bar{A}^{(3)}_{12}(\omega^\prime,\theta,\bar{q})&=&\cos\theta\;\bar{A}^{(3)}_{6}(
\omega^\prime,\theta,\bar{q})+\frac{g_{A}^2}{8\pi^2F_{\pi}^2}\int_{0}^{1}dx\int_{0}^{1}
dy\left\{\left(1-y\right)\left(2y-1\right)\frac{\omega^\prime\bar{q}\arcsin\left[
\frac{T}{\hat{m}}\right]}{\sqrt{\hat{m}^2-T^2}}\right. \nonumber \\
& &\phantom{\cos\theta\;\bar{A}^{(3)}_{4}(\omega^\prime,\theta,\bar{q}) }\left.
+2\cos\theta\left(1-x\right)x\left(1-y\right)^3y\;\frac{\bar{q}^2\omega^{\prime
2}\left(T\sqrt{\hat{m}^2-T^2}+\hat{m}^2\arcsin\left[\frac{T}{\hat{m}}\right]\right)}{
\hat{m}^2\left(\hat{m}^2-T^2\right)^{3/2}} \right\} \label{eq:a3a12}
\end{eqnarray}
Eqs.(\ref{eq:a1a2},\ref{eq:a3a12}) constitute the {\em full} ${\cal O}(p^3)$
HBChPT
results for the structure part in virtual Compton scattering off the nucleon.
As such,
they are independent of the particular formalism of Guichon and could also be
used
to extract alternative descriptions of generalized polarizabilities, {\em e.g.}
see
the recent paper by Unkmeir et al\cite{Christine}.

\subsubsection{Spin-independent Polarizabilities}\label{HBChPTscalar}

{From} the ${\cal O}(p^3)$ HBChPT results for the 12 structure amplitudes given
in the
previous section one can now extract the GPs as defined by Guichon, following
the general
formulae given in Eqs.(\ref{defab}-\ref{defcharge}). In this subsection we
first focus
on the spin-independent GPs $\bar{\alpha}_E(\bar{q}), \;
\bar{\beta}_M(\bar{q})$.

The leading $\bar{q}$-dependent modification of $\bar{\alpha}_E(\bar{q}), \;
\bar{\beta}_M(\bar{q})$ has already been analyzed in ref. \cite{HHKS1} and one
finds
\begin{eqnarray}
\bar{\alpha}^{(3)}_E(\bar{q}) &=& \frac{5 e^2 g_{A}^2}{384\pi^2 F_{\pi}^2
m_\pi}\left[1-\frac{7}{50}\frac{\bar{q}^2}{m_{\pi}^2}+\frac{81}{2800}\frac{
\bar{q}^4}{m_{\pi}^4}+{\cal O}(\bar{q}^6)\right] , \nonumber\\
\bar{\beta}^{(3)}_M(\bar{q}) &=& \frac{e^2 g_{A}^2}{768\pi^2 F_{\pi}^2
m_\pi}\left[1+\frac{1}{5}\frac{\bar{q}^2}{m_{\pi}^2}-\frac{39}{560}\frac{\bar{q}^4}{
m_{\pi}^4}+{\cal O}(\bar{q}^6)\right] . \label{abseries}
\end{eqnarray}
First, we note that in the limit $\bar{q}\rightarrow 0$ one recovers the
well-known
real Compton results at $q^2=0$ \cite{BKKM1}
\begin{eqnarray}
\bar{\alpha}^{(3)}_E &=& \bar{\alpha}^{(3)}_E(\bar{q}=0)=
\frac{5 e^2 g_{A}^2}{384\pi^2 F_{\pi}^2m_\pi}=12.5\times 10^{-4}\;\mbox{fm}^3
\nonumber \\
\bar{\beta}^{(3)}_M &=& \bar{\beta}^{(3)}_M(\bar{q}=0)=
\frac{e^2 g_{A}^2}{768\pi^2 F_{\pi}^2m_\pi}=1.25\times 10^{-4}\;\mbox{fm}^3 \; ,
\end{eqnarray}
which work extremely well when compared with the existing experimental
information
given in Eq.(\ref{abexp}).

As already pointed out in ref. \cite{HHKS1}, the slope of
$\bar{\alpha}_E(\bar{q}),
\; \bar{\beta}_M(\bar{q})$ with respect to $\bar{q}$ shows the {\it opposite
sign}
for the two spin-independent polarizabilities. These respective slopes are
uniquely determined by the chiral structure of the nucleon, i.e. the
``pion-cloud'', as given by the $\pi N$-loop diagrams of Fig.(\ref{figgnpi}).
At ${\cal
O}(p^3)$ ChPT therefore leads to the remarkable prediction that the
(generalized) magnetic polarizability $\bar{\beta}_M(\bar{q})$ {\em rises} with
increasing three-momentum transfer in a small window near $\bar{q}=0$. The
subleading, i.e. ${\cal O}(p^4)$, correction to this result is not known at
this point, but in section \ref{sres} we discuss the leading modification of
the slopes due to the $\Delta$(1232) resonance.

Starting from the expression for the individual Feynman diagrams given in
appendix \ref{appnpi},
the $\pi N$-loop contributions to
$\bar{\alpha}_E(\bar{q}),
\; \bar{\beta}_M(\bar{q})$ can be shown to possess
analytic expressions for their $\bar{q}$-dependence.  To ${\cal O}(p^3)$ we 
find the remarkably simple closed form expressions
\begin{eqnarray}
\bar{\alpha}^{(3)}_E(\bar{q})&=& \frac{e^2 g_{A}^2 m_\pi}{64\pi^2
F_{\pi}^2}\;\frac{4+2\frac{\bar{q}^2}{m_{\pi}^2}-\left(8-2\frac{\bar{q}^2}{m_{\pi}^2}
-\frac{\bar{q}^4}{m_{\pi}^4}\right)\frac{m_\pi}{\bar{q}}\arctan\frac{\bar{q}}{2
m_{\pi}}}{\bar{q}^2\left(4+\frac{\bar{q}^2}{m_{\pi}^2}\right)}
\; , \nonumber\\
\bar{\beta}^{(3)}_M(\bar{q})&=& \frac{e^2 g_{A}^2 m_\pi}{128\pi^2
F_{\pi}^2}\;\frac{-\left(4+2\frac{\bar{q}^2}{m_{\pi}^2}\right)+\left(8+6\frac{
\bar{q}^2}{m_{\pi}^2}+\frac{\bar{q}^4}{m_{\pi}^4}\right)\frac{m_\pi}{\bar{q}}\arctan
\frac{\bar{q}}{2 m_{\pi}}}{\bar{q}^2\left(4+\frac{\bar{q}^2}{m_{\pi}^2}\right)}
\; . \label{eq:bq}
\end{eqnarray}
These HBChPT predictions for
$\bar{\alpha}_E(\bar{q}), \; \bar{\beta}_M(\bar{q})$ are also shown in
Fig.\ref{figab}. One observes a relatively sharp fall-off in the electric GP,
whereas
the magnetic GP shows the {\em rising} behaviour for low values of $\bar{q}$ as
described
above. This remarkable
effect has its origin in the chiral structure of the pion cloud surrounding the
nucleon
and poses a formidable challenge to form-factor-supplemented Born-models of the
GPs
({\it e.g.} see \cite{Vanderhaeghen}). From
Eq.(\ref{eq:bq}) the maximum of the magnetic GP can be determined to be
\begin{equation}
\bar{\beta}_{M}^{max.}\left(\bar{q}=2.38 m_\pi\right)=1.29 \times
\bar{\beta}_M(0) \; ,
\end{equation}
indicating a 30\% enhancement of this GP relative to its value at the real
photon point.  Using the $C$-invariance relations
Eqs.(\ref{cinv}), we can also read off the remaining spin-independent GP
\begin{eqnarray}
\hat{P}^{(01,1)0}(\bar{q})&=&-{g_A^2m_\pi\over 16\pi F_\pi^2}{\omega_0\over
\bar{q}^4}
{4+2{\bar{q}^2\over m_\pi^2}+(-8+10{\bar{q}^2\over m_\pi^2}+3{\bar{q}^4\over
m_\pi^4})
{m_\pi\over \bar{q}}\arctan{\bar{q}\over 2m_\pi}\over (4+{\bar{q}^2\over
m_\pi^2})}\nonumber\\
&=& -{11g_A^2\over 576\pi F_\pi^2 M_Nm_\pi}\left[1-{6\over 55}
{\bar{q}^2\over m_\pi^2}+{123\over 560}{\bar{q}^4\over m_\pi^4}+{\cal
O}(\bar{q}^6)\right] \label{extra}
\end{eqnarray}
Once more we note that $\hat{P}^{(01,1)0}(\bar{q})$ is not an independent GP,
but
can be found as a linear combination of
$\bar{\alpha}^{(3)}_E(\bar{q}),\;\bar{\beta}^{(3)}_M(\bar{q})$ via the
charge-conjugation
constraint Eq.(\ref{cinv}). The extra suppression by $1/M_N$ compared to
Eq.(\ref{abseries}) arises from the expansion of the $\omega_0$ factor defined
in
Eq.(\ref{om0}).

Having discussed the scalar (spin-independent) structure
of the nucleon, we now move on to the spin-dependent analysis.

\subsubsection{Spin-dependent Generalized Polarizabilities}\label{HBChPTspin}

Following the identification of the GPs from the 12 structure amplitudes
via Eqs.(\ref{defspin}-\ref{defcharge}) we can also analyze the behaviour of
the
spin-dependent GPs near $\bar{q}=0$. For the four independent spin GPs we find
\begin{eqnarray}
P_{(01,12)1}^{(3)}(\bar{q})&=&-\frac{\sqrt{2}\;g_{A}^2}{288\pi^2F_{\pi}^2
                              m_{\pi}^2}\left[1-\frac{\bar{q}^2}{5\;m_{\pi}^2}+\frac{
                              3\;\bar{q}^4}{70\;m_{\pi}^4}+{\cal
                              O}(\bar{q}^6)\right] \nonumber\\
P_{(11,02)1}^{(3)}(\bar{q})&=&-\frac{\sqrt{2}\;g_{A}^2}{144\sqrt{3}\;\pi^2F_{\pi}^2
                              m_{\pi}^2}\left[1-\frac{\bar{q}^2}{5\;m_{\pi}^2}+\frac{3\;
                              \bar{q}^4}{70\;m_{\pi}^4}+{\cal
                              O}(\bar{q}^6)\right] \nonumber\\
P_{(11,00)1}^{(3)}(\bar{q})&=&-\frac{5\;g_{A}^2}{144\sqrt{3}\pi^2F_{\pi}^2}\left[0+
                              \frac{\bar{q}^2}{m_{\pi}^2}-\frac{7\;\bar{q}^4}{50\;
                              m_{\pi}^4}+{\cal O}(\bar{q}^6)\right] \nonumber\\
\hat{P}_{(01,1)1}^{(3)}(\bar{q})&=&-\frac{g_{A}^2}{48\sqrt{6}\;\pi^2F_{\pi}^2m_{\pi}^2}
                                    \left[1-\frac{2\;\bar{q}^2}{15\;m_{\pi}^2}+\frac{
                                    \bar{q}^4}{42\;m_{\pi}^4}+{\cal
                                    O}(\bar{q}^6)\right] , \label{spintaylor}
\end{eqnarray}
whereas the remaining three spin-dependent GPscan be determined via
Eq.(\ref{defcharge})
as a consequence of the $C$-invariance relations Eq.(\ref{cinv}):
\begin{eqnarray}
P_{(01,01)1}^{\;l.o.\;\chi}(\bar{q})&=&\frac{g_{A}^2}{144\pi^2F_{\pi}^2}\;\frac{1}{M_N}
                              \left[0
                              +\frac{\bar{q}^2}{m_{\pi}^2}-\left(\frac{3}{20}+\frac{
                              \mu^2}{4}\right)\frac{\bar{q}^4}{m_{\pi}^4}+{\cal
O}(
                              \bar{q}^6)\right] \nonumber\\
P_{(11,11)1}^{\;l.o.\;\chi}(\bar{q})&=&-\frac{g_{A}^2}{288\pi^2F_{\pi}^2}\;\frac{1}{M_N}
                              \left[0
                              +\frac{\bar{q}^2}{m_{\pi}^2}-\left(\frac{1}{10}-\frac{
                              \mu^2}{4}\right)\frac{\bar{q}^4}{m_{\pi}^4}+{\cal
O}(
                              \bar{q}^6)\right] \nonumber\\
\hat{P}_{(11,2)1}^{\;l.o.\;\chi}(\bar{q})&=&\frac{\sqrt{2}\;g_{A}^2}{2880\sqrt{5}\;\pi^2
                                   F_{\pi}^2m_{\pi}^2}\;\frac{1}{M_N}\left[0+\left(1+5\;
                                   \mu^2\right)\frac{\bar{q}^2}{m_{\pi}^2}-\left(\frac{
                                   2}{7}+\mu^2+\frac{15\;\mu^4}{4}\right)\frac{
                                   \bar{q}^4}{m_{\pi}^4}+{\cal
                                   O}(\bar{q}^6)\right] ,
\end{eqnarray}
with $\mu=m_\pi /M_N$. As in the case of $\hat{P}^{(01,1)0}(\bar{q})$ of
Eq.(\ref{extra}), one can clearly see that these three GPs are
formally suppressed
by an additional factor of $1/M_N$ relative to the four independent spin GPs of
Eq.(\ref{spintaylor}) and therefore ordinarily would not be accessible in a
${\cal
O}(p^3)$
calculation. It is only the charge-conjugation constraint that allows us to
extract them from the $\bar{A}_i^{(3)}$ VCS amplitudes.
It is also interesting to note that four of the generalized
spin-polarizabilities
{\it vanish} in the real Compton limit---$\bar{q}\rightarrow 0$. In the case
of $P^{(3)}_{(11,00)1}(\bar{q}), P^{l.o.\chi}_{(01,01)1}(\bar{q}),
P^{l.o.\chi}_{(11,11)1}(\bar{q})$ this follows
from charge conjugation invariance and crossing symmetry, as pointed out by
Drechsel et al.\cite{fm2}.  On the other hand, for
$\hat{P}^{l.o.\chi}_{(11,2)1}(\bar{q})$ the zero
appears to be a numerical accident which is only true at this order, since the
linear sigma model calculation of ref. \cite{Metz} violates this
condition. Nevertheless the zero
in the first three cases is a powerful confirmation of the
internal consistency of the ChPT approach to generalized polarizabilities.

As in the case of the spin-independent sector it is possible to give analytic
expressions for the 7 spin-dependent GPs. Defining the auxiliary function
\begin{equation}\label{auxfunction}
g(x)={{\rm sinh}^{-1}(x)\over x\sqrt{1+x^2}}\; ,
\end{equation}
the four independent generalized spin-polarizabilities to third order in the
chiral expansion read
\begin{eqnarray}
P_{(01,12)1}^{(3)}(\bar{q})&=& -\frac{g_{A}^2}{24\sqrt{2}\;\pi^2
                              F_{\pi}^2\bar{q}^2}\left[1-g({\bar{q}\over
2m_\pi})
                              \right]\nonumber\\
P_{(11,02)1}^{(3)}(\bar{q})&=&-\frac{g_{A}^2}{12\sqrt{6}\;\pi^2F_{\pi}^2\bar{q}^2}
                              \left[1-g({\bar{q}\over
2m_\pi})\right]\nonumber\\
P_{(11,00)1}^{(3)}(\bar{q})&=&\frac{g_{A}^2}{12\sqrt{3}\;\pi^2F_{\pi}^2}
                              \left[2-\left(2+{3\bar{q}^2\over 4m_\pi^2}\right)
                              g({\bar{q}\over
                              2m_\pi})\right] \nonumber\\
\hat{P}_{(01,1)1}^{(3)}(\bar{q})&=&\frac{g_{A}^2}{24\sqrt{6}\;\pi^2F_{\pi}^2\bar{q}^2}
                              \left[3-\left(3+{\bar{q}^2\over m_\pi^2}\right)
                              g({\bar{q}\over
                              2m_\pi})\right] .\label{4spin}
\end{eqnarray}
The ${\cal O}(p^3)$ HBChPT results for these four spin-dependent GPs are shown
in
Fig.(\ref{figspin1}). All are found to be negative in the low energy regime and
three
of them show a steep rise with
$\bar{q}$ at low three-momentum transfer---except for
$P_{(11,00)1}^{(3)}(\bar{q})$,
which vanishes for $\bar{q}\rightarrow 0$ and is strongly falling off for small
finite
values of $\bar{q}$. The remaining three C-constrained GPs are found to be
\begin{eqnarray}
P_{(01,01)1}^{\;l.o.\;\chi}(\bar{q})&=&\frac{g_{A}^2\omega_0}{24\pi^2F_{\pi}^2\bar{q}^2}
                        \left[1-(1+{\bar{q}^2\over 2 m_\pi^2})g({\bar{q}\over
                        2m_\pi})\right] ,\nonumber\\
P_{(11,11)1}^{\;l.o.\;\chi}(\bar{q})&=&-\frac{g_{A}^2\omega_0M_{N}^2}{24\pi^2
                                 F_{\pi}^2\bar{q}^4}\left[\left(\frac{2\;
                                 \omega_0}{M_N}+\frac{3\;\bar{q}^2}{M_{N}^2}\right)
                                 -\left(\frac{3\;m_{\pi}^2+
                                 \bar{q}^2}{M_{N}^2}\;\frac{\bar{q}^2}{m_{\pi}^2}+
                                 \frac{\omega_0}{M_N}\left(2+\frac{
                                 \bar{q}^2}{m_{\pi}^2}\right)\right)
                                 g({\bar{q}\over 2m\pi})\right],
                                 \nonumber\\
\hat{P}_{(11,2)1}^{\;l.o.\;\chi}(\bar{q})&=&\frac{g_{A}^2\omega_0M_{N}^2}{6\sqrt{10}\;
                                      \pi^2
                                      F_{\pi}^2\bar{q}^6}\;\left[
                                      \left(\frac{\omega_0}{M_N}+\frac{2\;
                                      \bar{q}^2}{M_{N}^2}\right)-{1\over 2}
                                      \left(\frac{\omega_0}{M_N}\left(2+\frac{
                                      \bar{q}^2}{m_{\pi}^2}\right)+\left(4+\frac{
                                      \bar{q}^2}{m_{\pi}^2}\right)\frac{\bar{q}^2}{
                                      M_{N}^2}\right)g({\bar{q}\over
                                      2m\pi})\right], \label{3spin}
\end{eqnarray}
with $\omega_0$ defined in Eq.(\ref{om0}). Their resulting
$\bar{q}$-dependence is shown in Fig.(\ref{figspin2}).
$P_{(01,01)1}(\bar{q}),\;
P_{(11,11)1}(\bar{q})$ vanish for $\bar{q}\rightarrow 0$ as required by
C-invariance
\cite{fm2}, whereas the unconstrained spin-dependent GP
$\hat{P}_{(11,2)1}(\bar{q})$
rises at low $\bar{q}$ and shows an unusual turnover point near
$\bar{q}^2\sim 0.2$ GeV$^2$. Once more we note that these three particular GPs,
strictly
speaking, lie beyond a ${\cal O}(p^3)$ calculation and could only be deduced via
the
C-invariance constraints of Eq.(\ref{cinv}).

{From} an analysis of the corresponding
spin-polarizabilities in real Compton scattering \cite{HHKK} one knows that
in some cases there exist
large corrections at $q^2=0$ to these chiral ${\cal O}(p^3)$ results of the
spin-polarizabilities due to the
Delta resonance. On the other hand, the spin-independent polarizabilities
$\bar{\alpha}_E,\;\bar{\beta}_M$ are known to be well described within
${\cal O}(p^3)$ $\pi N$ HBChPT (see Eq.(\ref{abseries}) in the limit
$\bar{q}\rightarrow 0$ and \cite{BKKM1}). In the next section we will therefore
analyze the leading effects of the $\Delta$(1232) on the GPs
in a {\em different chiral effective framework},
which contains the $\Delta$(1232) as an explicit degree of freedom.

\subsection{${\cal O}(\epsilon^3)$ Small Scale Expansion} \label{sres}
\subsubsection{General comments regarding SSE and Compton scattering}

In HBChPT the effects of $\Delta$(1232) are incorporated via higher order
contact
interactions, i.e. the effects of this particular resonance are not directly
tractable
in the calculation. If one is interested in such kind of questions, one needs a
chiral
effective framework which includes $\Delta$(1232) as an explicit degree of
freedom
{\it in a consistent power counting framework}---one approach of this kind is
SSE as
laid out in section \ref{deltatheory}.

First, we would like to stress again that any SSE calculation to ${\cal
O}(\epsilon^3)$
does not just equal the corresponding ${\cal O}(p^3)$ HBChPT calculation plus
some
additional diagrams with explicit Delta degrees of freedom. SSE constitutes a
chiral effective theory separate from HBChPT---for example,
even single nucleon coupling structures
which look the same in the (bare) lagrangians of the two theories can undergo
quite a
different coupling constant renormalization or acquire different
beta-functions, for
details we refer the interested reader to ref.\cite{bfhm}. For the particular
case
of (real) Compton scattering we would like to remind the reader that HBChPT
and
SSE show quite a different convergence behaviour for the (real) Compton
polarizabilities \cite{HHKK},
which is expected to also hold true for the here discussed generalized
polarizabilities
of VCS.

In principle there are two kinds of
{\it additional} contributions to the ${\cal O}(p^3)$ HBChPT results presented
in the
previous section---$\Delta$(1232) pole graphs (Fig.\ref{figgborn}(d,e)) and
$\Delta\pi$-continuum effects (Fig.\ref{figgdpi}).
The latter are straightforwardly obtained from the results given in
appendix \ref{appdpi}, whereas the Delta pole effects to be discussed here are identical
to their (real) Compton contributions discussed in \cite{HHKK}.

\subsubsection{Spin-independent results}\label{sir}

First we discuss the ${\cal O}(\epsilon^3)$ SSE results for the
spin-independent
GPs $\bar{\alpha}_E^{(III)}(\bar{q}),\;\bar{\beta}_M^{(III)}(\bar{q})$ near
$\bar{q}=0$
to facilitate the comparison between HBChPT and SSE. One finds
\begin{eqnarray}
\bar{\alpha}^{(III)}_E(\bar{q}) &=& \frac{5 e^2 g_{A}^2}{384\pi^2F_{\pi}^2m_\pi}
            +{e^2g_{\pi N\Delta}^2\over216\pi^3F_\pi^2}\left(
            {9\Delta\over \Delta^2-m_\pi^2}+{\Delta^2-10m_\pi^2\over
            (\Delta^2-m_\pi^2)^{3\over 2}}\ln R\right)
            \nonumber\\
         &+&\frac{\bar{q}^2}{m_\pi^2}\left[-\frac{7e^2g_{A}^2}{3840\pi^2
            F_{\pi}^2m_\pi}-{e^2g_{\pi N\Delta}^2\over1080\pi^3F_\pi^2}
            \left({2\Delta^3-17\Delta m_\pi^2\over
            (\Delta^2-m_\pi^2)^2}+{8\Delta^2m_\pi^2+7m_\pi^4\over (\Delta^2-
            m_\pi^2)^{5\over 2}}\ln R\right)\right]+{\cal O}(\bar{q}^4)
            \nonumber \\
         &=&\left\{12.5+4.22+\frac{\bar{q}^2}{m_\pi^2}\left[-1.75-0.240\right]
            +\frac{\bar{q}^4}{m_\pi^4}\left[0.362+0.018\right]+\dots\right\}
            \times 10^{-4}
            \;\mbox{fm}^3 \nonumber \\
\bar{\beta}^{(III)}_M(\bar{q}) &=& \frac{e^2 g_{A}^2}{768\pi^2 F_{\pi}^2m_\pi}
            +\frac{e^22\,b_1^2}{9\pi M_N^2\Delta}
            +{e^2g_{\pi N\Delta}^2\over 216\pi^3F_\pi^2}
            {1\over \sqrt{\Delta^2-m_\pi^2}}\ln R
            \nonumber\\
         &+&\frac{\bar{q}^2}{m_\pi^2}\left[\frac{e^2g_{A}^2}{3840\pi^2
            F_{\pi}^2m_\pi}+{e^2g_{\pi N\Delta}^2\over 1080\pi^3F_\pi^2}
            \left({\Delta\over (\Delta^2-m_\pi^2)}-{m_\pi^2\over (\Delta^2
            -m_\pi^2)^{3\over2}}\ln R\right)\right]+{\cal O}(\bar{q}^4)
            \nonumber \\
         &= &\left\{1.25+7.20+0.725+\frac{\bar{q}^2}{m_\pi^2}\left[0.250+0.078\right]
            +\frac{\bar{q}^4}{m_\pi^4}\left[-0.087-0.020\right]+\dots\right\}
            \times 10^{-4}\;\mbox{fm}^3 \label{SSEtaylorab}
\end{eqnarray}
with
\begin{equation}
R={\Delta\over m_\pi}+\sqrt{{\Delta^2\over m_\pi^2}-1}\; .
\end{equation}
The important point to note in Eq.(\ref{SSEtaylorab}) is the fact that the
$\bar{q}$-dependence is only modified in a very weak fashion by the inclusion 
of explicit delta degrees of freedom. In that respect SSE to ${\cal O}(\epsilon^3)$
and HBChPT to ${\cal O}(p^3)$ are quite compatible. However, the same problems known
from real Compton scattering \cite{delta,HHKK}
appear in the limit $\bar{q}\rightarrow 0$, which in the Guichon definition of the
GPs corresponds to the real photon point. In $\bar{\alpha}_E(0)\rightarrow 
\bar{\alpha}_E$ the 
$\Delta\pi$-continuum of Fig.(\ref{figgdpi}) 
produces a shift of $4.2\times 10^{-4}\;\mbox{fm}^3$, which when 
added to the $12.5\times 10^{-4}\;\mbox{fm}^3$ from the $N\pi$-continuum 
of Fig.(\ref{figgnpi}), leads to a much larger 
number then the current values of $\bar{\alpha}_E$ 
\cite{PPol}. In $\bar{\beta}_M(0)\rightarrow\bar{\beta}_M$ the effect 
is even more dramatic. Here it is 
the large magnetic contribution\footnote{As expected, $\bar{\alpha}_E$
is completely free of delta pole contributions to this order, quite analogous to the
case of real Compton scattering \cite{delta,HHKK}.} of $7.2\times 10^{-4}\;\mbox{fm}^3$
coming from the {\em $\bar{q}$-independent} delta pole graphs of 
Fig.(\ref{figgborn}d,e) which spoil any agreement
with the currently accepted number for $\bar{\beta}_M$ of the proton \cite{PPol}.
On the other hand, the sum of the contributions from the $N\pi$- and from the 
$\Delta\pi$-continuum has the right magnitude of 
$\sim\,2\times 10^{-4}\;\mbox{fm}^3$ for the magnetic polarizability, constituting
the ``chiral version'' of the unwanted presence of a large $\Delta$(1232)-induced
paramagnetism, which is well-known in the literature \cite{MNZ}. A large source of
diamagnetism due to the pion-cloud has been identified in refs.\cite{BKSM} in the
case of (real) Compton scattering, but 
this mechanism, which leads to a sensible (central) value of 
$\bar{\beta}_M\sim 3.5\;10^{-4}\,\mbox{fm}^3$ for the proton, can only
be implemented in a ${\cal O}(p^4)$ HBChPT (respectively ${\cal O}(\epsilon^4)$ SSE ?)
calculation and is therefore beyond the scope of this analysis. 
 
Keeping
these problems in mind, we nevertheless are convinced that the $\bar{q}$-dependence is 
described reasonably well 
by the ${\cal O}(\epsilon^3)$ calculation and that the problems 
described above only refer to the correct normalization of the theory at the real
photon point $\bar{q}\rightarrow 0$. We base this expectation on the observation that
the relevant scale of the $\bar{q}$-evolution in Eq.(\ref{SSEtaylorab})
at small momentum transfer is given by the quantity $\bar{q}^2/m_\pi^2$, i.e. the 
momentum dependence arises from the ``pion-cloud'' of the nucleon.
At the next order---${\cal O}(\epsilon^4)$---new diagrams are expected to correct the
normalization at the photon point. The $\bar{q}$-dependence of these diagrams however
is then expected to scale with $\bar{q}^2/(M_N m_\pi)$, i.e. it should be much weaker
due to the appearance of the extra suppression factor $m_\pi/M_N$ of the next order.
Whether this expectation will hold true 
can, of course, only be decided once 
$\bar{\alpha}_E(\bar{q}),\;\bar{\beta}_M(\bar{q})$ have been explictly calculated to 
${\cal O}(\epsilon^4)$. An analysis of the renormalization of $\bar{\alpha}_E,\;
\bar{\beta}_M$ in real Compton scattering to ${\cal O}(\epsilon^4)$ is under way 
\cite{GHKM} and
will later be extended to the case of VCS at ${\cal O}(\epsilon^4)$.

For completeness we also give formal expressions for the two spin-independent GPs. 
Unlike the case of ${\cal O}(p^3)$ HBChPT in SSE to ${\cal O}(\epsilon^3)$ 
we were not able to obtain
closed form expressions---
\begin{eqnarray}\label{eq:bqSSE}
\bar{\alpha}^{(III)}_E(\bar{q})&=& \frac{e^2 g_{A}^2 m_\pi}{64\pi^2
            F_{\pi}^2}\;\frac{4+2\frac{\bar{q}^2}{m_{\pi}^2}-
            \left(8-2\frac{\bar{q}^2}{m_{\pi}^2}
            -\frac{\bar{q}^4}{m_{\pi}^4}\right)\frac{m_\pi}{\bar{q}}
            \arctan\frac{\bar{q}}{2
            m_{\pi}}}{\bar{q}^2\left(4+\frac{\bar{q}^2}{m_{\pi}^2}\right)}
            \nonumber \\
         &+&\frac{e^2}{8\pi}\,\frac{8\,g_{\pi N\Delta}^2}{9\,F_\pi^2}
            \int_0^1 dx\int_0^1 dy\;\frac{\partial^2}{\partial w\,^2}\;
            \mbox{\Large \{}\;
            J_0\left(\omega^\prime-\Delta,m_\pi^2\right)+J_0\left(-\omega^\prime
            -\Delta,m_\pi^2\right)\nonumber \\
         & &-2\left[J_2^\prime\left(\omega^\prime x-\Delta,m_\pi^2\right)+J_2^\prime
                    \left(-\omega^\prime x-\Delta,m_\pi^2\right)+J_2^\prime
                    \left(\omega^\prime x-\Delta,\tilde{m}^2\right)+J_2^\prime
                    \left(-\omega^\prime x-\Delta,\tilde{m}^2\right)\right]
            \nonumber \\
         & &+4\left(1-y\right)\left[5\left(J_6^{\prime\prime}\left(T-\Delta,
                              \hat{m}^2\right)+J_6^{\prime\prime}\left(-T-\Delta,
                              \hat{m}^2\right)\right)-\left(T^2+m_\pi^2-\hat{m}^2-
                              T\,\omega^\prime\right)\right.\nonumber \\
         & &\phantom{+4\left(1-y\right)}\left. \left(J_2^{\prime\prime}\left(T-\Delta,
                              \hat{m}^2\right)+J_2^{\prime\prime}\left(-T-\Delta,
                              \hat{m}^2\right)\right)\right]
            \nonumber \\
         & &-2\left[3\,J_2^\prime\left(-\Delta,m_f^2\right)+\left(m_f^2-m_\pi^2\right)
                    J_0^\prime\left(-\Delta,m_f^2\right)\right]
            \nonumber \\
         & &-x\left(1-2x\right)\bar{q}^2\left(J_0^\prime
                    \left(\omega^\prime x-\Delta,\tilde{m}^2\right)+J_0^\prime
                    \left(-\omega^\prime x-\Delta,\tilde{m}^2\right)\right)
            \nonumber \\
         & &-2\left[\left(1-y\right)\left(14y^2-9y+1\right)\bar{q}^2\left(
                    J_2^{\prime\prime}\left(T-\Delta,\hat{m}^2\right)+
                    J_2^{\prime\prime}\left(-T-\Delta,\hat{m}^2\right)\right)\right.
                    \nonumber \\
         & &\phantom{-2}\left.+
                    y\left(1-y\right)\left(1-2y\right)\bar{q}^2\left(T^2+m_\pi^2
                    -\hat{m}^2-T\,w\right)\left(
                    J_0^{\prime\prime}\left(T-\Delta,\hat{m}^2\right)+
                    J_0^{\prime\prime}\left(-T-\Delta,\hat{m}^2\right)\right)\right]
            \mbox{\Large \} }\biggl|_{\,\cos\theta\rightarrow 0,
            \omega^\prime\rightarrow 0}\; , \nonumber\\
\bar{\beta}^{(III)}_M(\bar{q})&=& \frac{e^2 g_{A}^2 m_\pi}{128\pi^2
            F_{\pi}^2}\;\frac{-\left(4+2\frac{\bar{q}^2}{m_{\pi}^2}\right)
            +\left(8+6\frac{\bar{q}^2}{m_{\pi}^2}+\frac{\bar{q}^4}{m_{\pi}^4}
            \right)\frac{m_\pi}{\bar{q}}\arctan
            \frac{\bar{q}}{2 m_{\pi}}}{\bar{q}^2\left(4+\frac{\bar{q}^2}{
            m_{\pi}^2}\right)}+\frac{e^22\,b_1^2}{9\pi M_N^2\Delta}\nonumber\\
         &+&\frac{e^2}{4\pi}\,\frac{1}{\bar{q}}\,\frac{32\,g_{\pi N\Delta}^2}{9\,F_\pi^2}
            \int_0^1 dx\int_0^1 dy\;\frac{\partial}{\partial w}\;
            \mbox{\Large \{}\;\nonumber \\
         & &\left[\left(1-y\right)\left(-1+x-8xy+7(y-y^2+xy^2)\right)\bar{q}\,w
            \left(J_2^{\prime\prime}\left(T-\Delta,\hat{m}^2\right)+
            J_2^{\prime\prime}\left(-T-\Delta,\hat{m}^2\right)\right)\right.
            \nonumber \\
         & &\left.-y\left(1-y\right)^2\left(1-x\right)\left(T^2+m_\pi^2
            -\hat{m}^2-T\,w\right)\bar{q}\,w\left(J_0^{\prime\prime}
            \left(T-\Delta,\hat{m}^2\right)+
            J_0^{\prime\prime}\left(-T-\Delta,\hat{m}^2\right)\right)\right]
            \mbox{\Large \} }\biggl|_{\,\cos\theta\rightarrow 0,
            \omega^\prime\rightarrow 0} \; . \nonumber \\
\end{eqnarray}
We note that the relevant J-functions are defined in appendix \ref{Jfunctions} and the 
mass-/energy variables occuring in Eq.(\ref{eq:bqSSE}) have been given in
Eq.(\ref{definitions}).

The results of Eq.(\ref{eq:bqSSE}) are also shown in Fig.\ref{figabSSE}. Once more,
we do not advocate the use of these ${\cal O}(\epsilon^3)$ SSE curves in a realistic 
analysis of VCS at this point as there are large known cancellations which are not 
included yet to this order. A realistic use of these curves could be the 
prescription
\begin{eqnarray}
\bar{\alpha}_E^{ren.}(\bar{q})&=&\bar{\alpha}_E^{(III)}(\bar{q})-
                                 \bar{\alpha}_E^{(III)}(0)+\bar{\alpha}_E^{exp.}
                                 \nonumber \\
\bar{\beta}_M^{ren.}(\bar{q})&=&\bar{\beta}_M^{(III)}(\bar{q})
                                -\bar{\beta}_M^{(III)}(0)+\bar{\beta}_M^{exp.}\;,
                                \label{prescript}
\end{eqnarray}
where the index $exp.$ refers to the current experimental numbers for 
$\bar{\alpha}_E,\;\bar{\beta}_M$ of ref.\cite{PPol}. The results of this operation
are shown in Fig.\ref{figabren}. There one can clearly see that the $\Delta$(1232) 
related effects at ${\cal O}(\epsilon^3)$ SSE {\em enhance} the $\bar{q}$-trend already 
seen at ${\cal O}(p^3)$ HBChPT. Of course we want to emphasize that the prescription
of Eq.(\ref{prescript}) leaves the strict realm of chiral effective theories and just
constitutes an ad hoc fix to include {\em some} effects that are of higher order in 
the (slowly converging) SSE expansion for the spin-independent GPs. 

Now we move to the generalized spin polarizabilities in SSE.

\subsubsection{Spin-dependent results}

Once more we start from a discussion of the GPs near $\bar{q}=0$.
It should be noted that there are no $\Delta$ pole contributions\footnote{We
observe
that there does exist a $\Delta$-pole contribution to the spin GP
$\hat{P}_{(01,1)0}(\bar{q})$,
\begin{equation}
\hat{P}_{(01,1)0}^{\Delta-{\rm pole}}=-{4\omega_0\over 27\bar{q}^2}{b_1^2\over
M^2\Delta}\;,
\end{equation}
which, however, is suppressed by an additional factor of $1/M_N$ originating
in $\omega_0$ of Eq.(\ref{om0}) and therefore is counted as
a ${\cal O}(\epsilon^4)$ effect.}
to any of the
generalized spin polarizabilities at ${\cal O}(\epsilon^3)$, quite in contrast
to the real Compton (Ragusa) spin polarizabilities $\gamma_2,\;\gamma_4$
\cite{HHKK}.
The ${\cal O}(\epsilon^3)$ results for the four independent spin GPs therefore
exclusively arise from the $N\pi$- and $\Delta\pi$-continuum graphs of
Figs.(\ref{figgnpi},\ref{figgdpi}) and can be found from the expressions given
in appendices \ref{appnpi} and \ref{appdpi}. One obtains
\begin{eqnarray}\label{SSEspinreihe}
P_{(01,12)1}^{(III)}(\bar{q})&=&-\frac{\sqrt{2}\;g_{A}^2}{288\pi^2F_{\pi}^2m_{\pi}^2}
            -{\sqrt{2}g_{\pi N\Delta}^2\over 324\pi^2F_\pi^2}\left({1\over
            \Delta^2-m_\pi^2}-{\Delta\over (\Delta^2-m_\pi^2)^{3\over 2}}
            \ln R\right)\nonumber\\
         & &+\frac{\bar{q}^2}{m_\pi^2}\left[\frac{\sqrt{2}\;g_{A}^2}{1440
            \pi^2F_{\pi}^2m_{\pi}^2}-{\sqrt{2}g_{\pi N\Delta}^2\over
            3240\pi^2F_\pi^2}
            \left({\Delta^2+2m_\pi^2\over (\Delta^2-m_\pi^2)^2}
            -{3\Delta m_\pi^2\over (\Delta^2-m_\pi^2)^{5\over 2}}\ln R\right)
            \right] +{\cal O}(\bar{q}^4) \nonumber \\
         &=&\left\{-7.28+0.735+\frac{\bar{q}^2}{m_\pi^2}\left[1.46-0.067\right]
            +\frac{\bar{q}^4}{m_\pi^4}\left[-0.312+0.009\right]+\dots\right\} \times
            10^{-3}\;\mbox{fm}^4\nonumber \\
P_{(11,02)1}^{(III)}(\bar{q})&=&-\frac{\sqrt{2}\;g_{A}^2}{144\sqrt{3}\;\pi^2F_{\pi}^2
            m_{\pi}^2} -\sqrt{2\over 3}{g_{\pi N\Delta}^2\over 162\pi^2
            F_\pi^2}\left({1\over \Delta^2-m_\pi^2}-{\Delta\over
            (\Delta^2-m_\pi^2)^{3\over 2}}\ln R\right) \nonumber\\
         & &+\frac{\bar{q}^2}{m_\pi^2}\left[\frac{\sqrt{2}\;g_{A}^2}{720\sqrt{3}
            \;\pi^2F_{\pi}^2m_{\pi}^2}-\sqrt{2\over 3}{g_{\pi N\Delta}^2\over 1620\pi^2
            F_\pi^2}\left({\Delta^2+2m_\pi^2\over (\Delta^2-m_\pi^2)^2}
            -{3\Delta m_\pi^2\over (\Delta^2-m_\pi^2)^{5\over 2}}\ln R\right)
            \right] +{\cal O}(\bar{q}^4) \nonumber \\
         &=&\left\{-8.41+0.848+\frac{\bar{q}^2}{m_\pi^2}\left[1.68-0.077\right]
            +\frac{\bar{q}^4}{m_\pi^4}\left[-0.360+0.010\right]\dots\right\} \times
            10^{-3}\;\mbox{fm}^4\nonumber \\
P_{(11,00)1}^{(III)}(\bar{q})&=&0+\frac{\bar{q}^2}{m_{\pi}^2}
            \left[-\frac{5\;g_{A}^2}{144\sqrt{3}\pi^2F_{\pi}^2}
            -\sqrt{1\over 3}{5g_{\pi N\Delta}^2\over 162\pi^2
            F_\pi^2}\left({m_\pi^2\over \Delta^2-m_\pi^2}-{\Delta m_\pi^2\over
            (\Delta^2-m_\pi^2)^{3\over 2}}\ln R\right)\right]+{\cal O}(\bar{q}^4)
            \nonumber\\
         &=&\left\{0+0+\frac{\bar{q}^2}{m_\pi^2}\left[-1.49+0.15\right]
            +\frac{\bar{q}^4}{m_\pi^4}\left[0.208-0.002\right]+\dots\right\} \times
            10^{-2}\;\mbox{fm}^2 \nonumber\\
\hat{P}_{(01,1)1}^{(III)}(\bar{q})&=&-\frac{g_{A}^2}{48\sqrt{6}\;\pi^2
            F_{\pi}^2m_{\pi}^2}-\sqrt{1\over 6}{g_{\pi N\Delta}^2\over
            54\pi^2F_\pi^2}\left({1\over \Delta^2-m_\pi^2}-{\Delta\over
            (\Delta^2-m_\pi^2)^{3\over 2}}\ln R\right)\nonumber\\
         & &+\frac{\bar{q}^2}{m_\pi^2}\left[\frac{g_{A}^2}{360\sqrt{6}\;\pi^2
            F_{\pi}^2m_{\pi}^2}-\sqrt{1\over 6}{g_{\pi N\Delta}^2\over
            810\pi^2F_\pi^2}\left({\Delta^2+2m_\pi^2\over (\Delta^2-m_\pi^2)^2}
            -{3\Delta m_\pi^2\over (\Delta^2-m_\pi^2)^{5\over 2}}\ln R\right)
            \right] +{\cal O}(\bar{q}^4) \nonumber \\
         &=&\left\{-12.6+1.272+\frac{\bar{q}^2}{m_\pi^2}\left[1.68-0.077\right]
            +\frac{\bar{q}^4}{m_\pi^4}\left[-0.300+0.009\right]+\dots\right\} \times
            10^{-3}\;\mbox{fm}^4.
\end{eqnarray}
First, we observe that SSE to ${\cal O}(\epsilon^3)$ obeys the C-invariance
constraint
\cite{fm2} $\lim_{\bar{q}\rightarrow 0}P_{(11,00)1}^{(III)}(\bar{q})=0$, as
does the
${\cal O}(p^3)$ HBChPT calculation in Eq.(\ref{spintaylor}). Second, we note
that there is no strong renormalization of the above\footnote{This is to be 
contrasted with the individual (real) Compton spin polarizabilities 
$\gamma_2$ and $\gamma_4$
defined by Ragusa, see ref.\cite{HHKK}. To be more specific about the connection
between VCS and real Compton scattering we utilize Eq.(\ref{spinconnection})
and find
\begin{eqnarray}
\gamma_3^{(III)}&=&1.0\times 10^{-4}\;\mbox{fm}^4 \nonumber \\
\gamma_2^{(III)}+\gamma_4^{(III)}&=&1.0\times 10^{-4}\;\mbox{fm}^4\; ,
\end{eqnarray}
with the input from Eq.(\ref{SSEspinreihe}). This is in complete agreement with
the results of ref.\cite{HHKK}. It turns out that the large $\Delta$(1232) pole
contribution cancels in this particular linear combination of
$\gamma_2$ and $\gamma_4$.} spin-dependent GPs at the
real photon point due to $\Delta$(1232) related effects. We observe that in
general the effects from the $\Delta\pi$-continuum are small and {\it always}
interfere {\it destructively} with the corresponding contribution from the 
$N\pi$-continuum, in contrast to the {\it constructive} interference in the
spin-independent sector of section \ref{sir}.

As in the previous section, we were not able to give the full
spin-dependent
${\cal O}(\epsilon^3)$ results in a closed form expression but utilize a
Feynman-parameter representation and the J-functions defined in appendix 
\ref{Jfunctions}:
\begin{eqnarray}\label{sdspin}
P_{(01,12)1}^{(III)}(\bar{q})&=& -\frac{g_{A}^2}{24\sqrt{2}\;\pi^2
                              F_{\pi}^2\bar{q}^2}\left[1-g({\bar{q}\over
                              2m_\pi})
                              \right]\nonumber \\
   & &-\frac{\sqrt{2}}{3}\,\frac{1}{\bar{q}}\,\frac{16\,g_{\pi N\Delta}^2}{9\,F_\pi^2}
      \int_0^1 dx\int_0^1 dy\;\frac{\partial}{\partial w}\;\mbox{\Large \{}
      \left[y\left(1-y\right)\bar{q}^2\left(J_2^{\prime\prime}\left(T-\Delta,\hat{m}^2
      \right)-J_2^{\prime\prime}\left(-T-\Delta,\hat{m}^2\right)\right)\right.
      \nonumber \\
   & &\left. +x\left(1-x\right)y\left(1-y\right)^3\bar{q}^2\omega^{\prime\,2}\left(
      J_0^{\prime\prime}\left(T-\Delta,\hat{m}^2\right)-J_0^{\prime\prime}
      \left(-T-\Delta,\hat{m}^2\right)\right)\right]
      \mbox{\Large \} }\biggl|_{\,\cos\theta\rightarrow 0,\omega^\prime\rightarrow 0}
       \nonumber\\
P_{(11,02)1}^{(III)}(\bar{q})&=&-\frac{g_{A}^2}{12\sqrt{6}\;\pi^2F_{\pi}^2\bar{q}^2}
                              \left[1-g({\bar{q}\over 2m_\pi})\right]\nonumber \\
   &-&\frac{\sqrt{2}}{3\sqrt{3}}\,\frac{1}{\bar{q}}\,\frac{8\,g_{\pi N\Delta}^2}
      {9\,F_\pi^2} \int_0^1 dx\int_0^1 dy\;\frac{\partial^2}{\partial w^{\prime\;2}}
      \;\mbox{\Large \{}
      \left[x\,y\left(1-y\right)^2\left(1-2y\right)\bar{q}^3\omega^\prime\left(
      J_0^{\prime\prime}\left(T-\Delta,\hat{m}^2\right)-J_0^{\prime\prime}\left(
      -T-\Delta,\hat{m}^2\right)\right)\right] \nonumber\\
   & &+2\left[x\left(1-y\right)^2\bar{q}\,\omega^\prime\left(
      J_2^{\prime\prime}\left(T-\Delta,\hat{m}^2\right)-J_2^{\prime\prime}\left(
      -T-\Delta,\hat{m}^2\right)\right)\right] \nonumber\\
   & &-2\left[y\left(1-y\right)\bar{q}\,\omega^\prime\left(
      J_2^{\prime\prime}\left(T-\Delta,\hat{m}^2\right)-J_2^{\prime\prime}\left(
      -T-\Delta,\hat{m}^2\right)\right)\right]
      \mbox{\Large \} }\biggl|_{\,\cos\theta\rightarrow 0,\omega^\prime\rightarrow 0}
      \nonumber\\
P_{(11,00)1}^{(III)}(\bar{q})&=&\frac{g_{A}^2}{12\sqrt{3}\;\pi^2F_{\pi}^2}
                              \left[2-\left(2+{3\bar{q}^2\over 4m_\pi^2}\right)
                              g({\bar{q}\over 2m_\pi})\right]
     +\sqrt{2}\,\bar{q}^2\,P_{(11,02)1}^{(III)}(\bar{q})
     -\frac{\bar{q}}{\sqrt{3}}\,\frac{8\,g_{\pi N\Delta}^2}
      {9\,F_\pi^2} \int_0^1 dx\int_0^1 dy\;\frac{\partial^2}{\partial w^{\prime\;2}}
      \;\mbox{\Large \{}\nonumber\\
   & &\left[\left(1-y\right)\left(1-2y\right)\bar{q}\,\omega^\prime
      \left(J_2^{\prime\prime}\left(T-\Delta,\hat{m}^2\right)-J_2^{\prime\prime}
      \left(-T-\Delta,\hat{m}^2\right)\right)\right]
      \mbox{\Large \} }\biggl|_{\,\cos\theta\rightarrow 0,\omega^\prime\rightarrow 0}
      \nonumber\\
\hat{P}_{(01,1)1}^{(III)}(\bar{q})&=&\frac{g_{A}^2}{24\sqrt{6}\;\pi^2F_{\pi}^2\bar{q}^2}
                              \left[3-\left(3+{\bar{q}^2\over m_\pi^2}\right)
                              g({\bar{q}\over 2m_\pi})\right]
                              +\sqrt{1\over 3}P_{(01,12)1}^{(III)}(\bar{q})\nonumber\\
  &-&\frac{2\sqrt{2}}{3\sqrt{3}}\,\frac{1}{\bar{q}^2}\,\frac{4\,g_{\pi N\Delta}^2}
      {9\,F_\pi^2} 
      \int_0^1 dx\int_0^1 dy\;\frac{\partial}{\partial w^{\prime}}
      \;\mbox{\Large \{}\left[-\left(J_0^\prime\left(\omega^\prime-\Delta,m_\pi^2\right)
      -J_0^\prime\left(-\omega^\prime-\Delta,m_\pi^2\right)\right)\right]\nonumber\\
  & &+2\,\frac{3}{d-1}
     \left[J_2^\prime\left(\omega^\prime x-\Delta,m_\pi^2\right)-J_2^\prime
     \left(-\omega^\prime x-\Delta,m_\pi^2\right)+J_2^\prime\left(\omega^\prime x-
     \Delta,\tilde{m}^2\right)-J_2^\prime\left(-\omega^\prime x-\Delta,\tilde{m}^2
     \right)\right]\nonumber\\
  & &-4\left[x\left(1-x\right)y\left(1-y\right)^3\bar{q}^2\omega^{\prime\;2}\left(
     J_0^{\prime\prime}\left(T-\Delta,\hat{m}^2\right)-J_0^{\prime\prime}\left(
     -T-\Delta,\hat{m}^2\right)\right)\right]
     \mbox{\Large \} }\biggl|_{\,\cos\theta\rightarrow 0,\omega^\prime\rightarrow 0}
\end{eqnarray}
Note that the auxiliary function $g(x)$ has already been defined in 
Eq.(\ref{auxfunction}) and the mass-/energy-variables again correspond to the
structures introduced in Eq.(\ref{definitions}). We present the {\em absolute} 
${\cal O}(\epsilon^3)$ SSE predictions 
for the four independent spin GPs in Fig.\ref{figSSEspin1}. It clearly shows 
that the ${\cal O}(\epsilon^3)$ curves are always lying higher than the corresponding
${\cal O}(p^3)$ HBChPT ones. In all cases the two curves share a similar behaviour in
their $\bar{q}$-dependence---leading to the conclusion that there is ``no dramatic''
signal of the $\Delta$(1232) resonance in the spin-dependent GPs to 
${\cal O}(\epsilon^3)$ compared to the dominant contributions from the $N\pi$-continuum.

Finally we note that the remaining (linearly dependent) generalized spin 
polarizabilities $P_{(01,01)1}^{(III)},P_{(11,11)1}^{(III)},\hat{P}_{(11,2)1}^{(III)}$ 
may be found via the charge-conjugation
constraint Eq.(\ref{cinv}).

\section{The Mainz Experiment}

As mentioned above, the pioneering VCS experiment\footnote{We note
that the theoretical 
predictions of the $\bar{q}$-dependence in the GPs given in 
refs.\cite{HHKS1,HHKS2} 
{\it preceded} the analysis of the experiment at Mainz.} has taken
place at Mainz, and
preliminary results of the analysis are now available \cite{nstar}.  
The measurement was performed at
$\bar{q}^2=0.36$ GeV$^2$ and used parallel kinematics although relativistic
forward-focussing allowed access to events as much as $\pm$ 26 degrees out of
plane.
Nevertheless the desired generalized polarizabilities were hidden behind a very
large
Bethe-Heitler background and their extraction was a real experimental tour de
force. Consulting Fig.\ref{figab},
we note that at $\bar{q}^2=0.36$ GeV$^2$ the ${\cal O}(p^3)$ HBChPT calculation
predicts that $\bar{\alpha}_E(\bar{q})$ should have decreased by as much as
50\% from
its real photon value, whereas the much smaller GP
$\bar{\beta}_M(\bar{q})$ is predicted to have slightly increased. As
can be seen from the HBChPT predictions 
in Figs.\ref{figspin1},\ref{figspin2}, the
spin-dependent GPs will dramatically change with regard to 
the real photon point.  Thus the
confrontation of theoretical predictions with the MAMI results offers a chance
to realistically test theoretical pictures of nucleon structure.  
Essentially two quantities were determined experimentally---the 
combination $P_{LL}-P_{TT}/\epsilon$ 
of longitudinal and transverse
response functions, which is primarily sensitive to the generalized electric 
polarizability $\alpha_E(\bar{q})$ (plus linear combinations of spin GPs) 
\cite{nstar,Guichon}, as well as
the interference term $P_{LT}$, depending on the generalized magnetic 
polarizability $\beta_M(\bar{q})$ and the spin GP
$P^{(01,01)1}(\bar{q})$ \cite{nstar,Guichon} 
(which itself can be expressed as a linear combination 
of the 2 spin GPs $P^{(11,00)1}(\bar{q})$ and $P^{(11,02)1}(\bar{q})$ via the
C-invariance constraint of Eq.(\ref{cinv})).
Results of the experiment together with predictions from ${\cal O}(p^3)$ 
HBChPT\footnote{We do not give predictions for the response functions of 
${\cal O}(\epsilon^3)$ SSE due to the discussed normalization problem in 
$\bar{\alpha}_E^{(III)}(\bar{q}=0)$. However, we believe that the 
${\cal O}(\epsilon^3)$ SSE predictions for the spin GPs will be helpful 
for ongoing studies on double polarization VCS experiments, which might
provide the possibility to study the connection between Ragusa and Guichon
spin-polarizabilities as indicated by Eq.(\ref{spinconnection}).} and other 
theoretical models are given in Table 1.
Obviously only the chiral picture (refs.\cite{HHKS1,HHKS2}; sections 
\ref{HBChPTscalar}, \ref{HBChPTspin} of this work; Figs. \ref{figab}, \ref{figspin1},
\ref{figspin2})
is able to explain the experimental features at this point.
Of course this is
only a single experiment at a single momentum transfer and results from other
laboratories and other values of $\bar{q}^2$ are needed in order to confirm
our predictions.  Nevertheless the agreement is certainly encouraging.

\section{Summary}

Virtual Compton scattering---$eN\rightarrow
e'N\gamma$---opens the way to high resolution study of nucleon structure by
measuring generalized polarizabilities (GPs), which are
momentum-dependent analogues of the
familiar polarizabilities determined in real Compton scattering.  In this work,
we
have calculated these quantities within the framework of conventional
heavy baryon chiral perturbation theory to third order in the momentum
expansion as well as to third order in the ``small-scale
expansion'', which contains the $\Delta(1232)$ as
an explicit degree of freedom.  As originally defined by Guichon et al., there
exist ten such GPs, three being associated with spin-independent correlations
and seven connected with spin-flip structures.   At third order
both in HBChPT and in SSE only six of these---two spin-independent and four
spin-dependent---survive, and we have calculated these directly.
In the case of the $\Delta$-pole and $\pi N$ loop contributions, we were able to 
obtain results for the GPs which are simple analytic forms,
while in the case of the corrresponding $\pi\Delta(1232)$-continuum contributions 
only numerical results could be given. We briefly discussed the results from the
first VCS experiment on the proton from Mainz at $Q^2=0.33$ GeV$^2$.
The success in predicting the measured
response functions resulted from a combination of a sharp falloff 
of $\bar{\alpha}_E^{(3)}(\bar{q})$, a slight rise of
$\bar{\beta}_M^{(3)}(\bar{q})$ and a strong incarease 
in the contributing spin GPs with
momentum-transfer $\bar{q}$. All these effects are intimately related to the chiral 
dynamics of the pion cloud, which can be calculated very precisely in chiral effective
theories like HBChPT and SSE---with HBChPT at least in the spin-independent sector having
the better convergence behaviour as far as we can tell at this point. 
In particular, for the case of the
generalized magnetic polarizability both HBChPT and SSE predict a {\it rising} 
behaviour as one goes away from the real photon
point---$\bar{q}^2=0$---up to a momentum $\bar{q}^2\sim 0.1$
GeV$^2$. This is a distinctive feature of the chiral calculations and 
generally not found in simple quark model evaluations.
It expresses the feature that chiral invariance
requires local regions {\it both} of paramagnetic (at small distances)
and diamagnetic (at larger distances)
magnetic polarizability densities in the nucleon.
Aside from 
the widely discussed $\bar{q}$-dependences of the generalized electric and 
magnetic polarizabilities, the strong variation of the GPs in the spin-sector is likely 
to be of interest for further study, both on the experimental and on the theoretical
side. Considering the results of the chiral calculations for the spin
polarizabilities in real Compton scattering we believe that the
${\cal O}(\epsilon^3)$ SSE calculation should be quite competitive with
the ${\cal O}(p^3)$ HBChPT analysis at least as far as the generalized 
spin-polarizabilities are concerned. Future measurements at Bates, 
MAMI and TJNAF will clarify this issue.

It goes without saying that our calculation is preliminary in that
it does not include important corrections arising
at ${\cal O}(p^4)/{\cal O}(\epsilon^4)$---see, {\it e.g.} 
the discussed normalization 
problems in $\bar{\alpha}_E^{(III)}(0),\,\bar{\beta}_M^{(III)}(0)$.  
An ${\cal O}(p^4)$ HBChPT analysis has been carried out in the case of the
real Compton electric and magnetic polarizabilities in ref.\cite{BKSM}
and important corrections and
uncertainties were found which, while not drastically modifying the basic numerical
predictions obtained at ${\cal O}(p^3)$, did introduce sizable
uncertainties into the predictions due to unknown counterterms which had to
be estimated via resonance exchange.  We may then anticipate 
a similar behavior here---that such higher order corrections will 
not change the basic pattern of the chiral 
${\cal O}(p^3)/{\cal O}(\epsilon^3)$ predictions, but mainly 
only correct the (photon-point) 
normalization.  However, verification of this assumption awaits detailed
future calculations. Lastly, we stress once more the motivation for 
performing electron scattering experiments on the nucleon: Different 
theoretical approaches may yield comparable results at the real photon
point, but the details
of the underlying dynamics can be analyzed in a much more 
powerful way by studying the $Q^2$-dependence. In conclusion, VCS on the 
nucleon has matured to become a precise testing ground for our notions
of nucleon structure at low energies.

\bigskip

\begin{center}
{\bf Acknowledgments}
\end{center}
The authors acknowledge many helpful discussions with N. d'Hose, U.-G. Mei{\ss}ner,
A. Metz, R. Miskimen, S. Scherer, J. Shaw and M. Vanderhaeghen.
BRH would like to acknowledge the support of the Alexander von Humboldt
Foundation
and the National Science Foundation, as well as the
hospitality of the IKP at Forschungszentrum J{\" u}lich.  This work is also
supported in part by the Deutsche Forschungsgemeinschaft (SFB443).

\begin{appendix}
\section{Loop Functions}\label{Jfunctions}

The formalism to calculate the loop diagrams for Compton
scattering both in ChPT and in the small scale expansion has been described in
detail in the appendices of ref.\cite{delta}. Therefore we shall only
give some definitions of the basic building blocks.

We express the invariant amplitudes of Feynman diagrams containing
pion-nucleon loops in terms of $d$-dimensional J-functions, defined via
\begin{eqnarray}
{1\over i}\int{d^d\ell\over (2\pi)^d}
{\{1,\ell_\mu\ell_\nu,\ell_\mu\ell_\nu\ell_\alpha\ell_\beta\}\over
(v\cdot\ell-W-i\eta)(M^2-\ell^2-i\eta)}&=& \left\{J_0(W,M),
g_{\mu\nu}J_2(W,M)+v_\mu v_\nu J_3(W,M), \right.\nonumber \\
& &\left. (g_{\mu\nu} g_{\alpha\beta}+{\rm perm.})J_6(W,M)+\dots \right\} ,
\label{eq:props}
\end{eqnarray}
with the small imaginary part $\eta$ denoting the location of the pole.

In the case of Compton scattering at ${\cal O}(p^3)$ or ${\cal O}(\epsilon^3)$,
all loop-integrals can be expressed in terms of the four functions
$\Delta_M,J_0(W,M),J_2(W,M),J_6(W,M)$, which are related via
\begin{eqnarray}
J_2 \left(W,M\right) & = & \frac{1}{d-1} \left[ \left(
    M^2 - W^2 \right) J_0 \left(W,M\right)
    - W \; \Delta_{M} \right] \label{J_2}\nonumber\\
J_6 \left(W,M\right) & = & \frac{1}{d+1} \left[ \left(
    M^2 - W^2 \right) J_2 \left(W,M\right)
    - \frac{M^2 W}{d} \; \Delta_M \right] \label{J_6} \, ,
\end{eqnarray}
with $\Delta_M$ denoting the meson integral
\begin{equation}
\Delta_M=\frac{1}{i}\int\frac{d^d \ell}{(2\pi)^d}\;\frac{1}{M^2-\ell^2-i\eta}
\; ,
\end{equation}
and $J_0\left(W,M\right)$ being the basic meson-baryon integral with arbitrary energy
$W$ and mass variable $M$.
Explicit representations for these building blocks can be found in appendix
A of ref.\cite{delta}.

Finally, we remind the reader that
all propagator structures encountered in the calculation can be reduced to
the basic forms of Eq.(\ref{eq:props}) by taking derivatives of the J-functions
with respect to the square of the mass---
\begin{eqnarray}
J_i'\left(W,M\right) & = & \frac{\partial}{\partial
\left( M^2 \right)} J_i \left(W,M\right) \, , \nonumber\\
J_i''\left(W,M\right) & = & \frac{\partial^2}{\partial
\left( M^2 \right)^2} J_i \left(W,M\right) .
\end{eqnarray}
For a more detailed discussion we refer to ref.\cite{trh}.

\section{$N\pi$ Loop Amplitudes in VCS}\label{appnpi}

Using the J-function formalism defined in Appendix A, one can get exact
solutions for the nine $N\pi$-loop diagrams of Fig.\ref{figgnpi}.
By $\tilde{\epsilon}_\mu \;
(q_\mu)$ we denote the polarization-vector
(four-momentum) of the incoming virtual photon, and by 
$\epsilon_{\mu}^{\prime} \; (q^{\prime}_\mu)$ the corresponding
quantities in the outgoing real photon with energy $\omega^\prime$. In order
to make contact with the VCS amplitudes defined in Eq.(\ref{eq:vcs12}),
we use the Coulomb gauge 
\begin{equation}
\tilde{\epsilon}^\mu=(0,\vec{\epsilon}_T
+{q^2\over \omega^2}\vec{\epsilon}\cdot\hat{q}\hat{q})
\end{equation}
The amplitudes can then be cast in the form
\begin{eqnarray}
Amp_{1+2}^{N\pi}  &=& i \frac{g_{A}^2}{F_{\pi}^2} \; \bar{u}_2(r^\prime)
                      \left\{ -\frac{1}{2} \tilde{\epsilon} \cdot
\epsilon^\prime
                      \left[ J_0 ( \omega^\prime , m_{\pi}^2 )
+J_0(-\omega^\prime ,
                      m_{\pi}^2 ) \right] + \; [ S \cdot \epsilon^\prime,S\cdot
                      \tilde{\epsilon} ] \left[ J_0(\omega^\prime ,m_{\pi}^2)-
                      J_0(-\omega^\prime ,m_{\pi}^2)
                      \right] \right\} u_1(r)  \nonumber\\
Amp_{3+6}^{N\pi}  &=& i \frac{g_{A}^2}{F_{\pi}^2} \int_{0}^{1}dx \;
                      \bar{u}_2(r^\prime) \left\{ \tilde{\epsilon}
\cdot\epsilon^\prime
                      \left[ J^{\prime}_2 ( \omega^\prime x , m_{\pi}^2 ) +
                      J^{\prime}_2 ( -\omega^\prime x, m_{\pi}^2 )
                      \right] \right. \nonumber \\
                  & & \phantom{i \frac{g_{A}^2}{F_{\pi}^2} \int_{0}^{1}dx \;
                      \bar{u}_2(r^\prime) } \left.-2 \; [S\cdot\epsilon^\prime,
                      S\cdot \tilde{\epsilon} ] \left[ J^{\prime}_2
                      ( \omega^\prime x,m_{\pi}^2
                      ) - J^{\prime}_2 ( -\omega^\prime x, m_{\pi}^2 )
                      \right] \right\} u_1(r)  \nonumber\\
Amp_{4+5}^{N\pi}  &=& i \frac{g_{A}^2}{F_{\pi}^2} \int_{0}^{1}dx \;
                      \bar{u}_2(r^\prime) \left\{
                      \tilde{\epsilon} \cdot \epsilon^\prime
                      \left[ J_{2}^\prime ( \omega^\prime x , \tilde{m}^2 ) +
                       J_{2}^\prime ( - \omega^\prime x
,\tilde{m}^2)\right]\right.
                      \nonumber \\
                  & & \phantom{i \frac{g_{A}^2}{F_{\pi}^2} \int_{0}^{1}dx \;
                      \bar{u}_2(r^\prime) }
                      -2 \; [ S\cdot \epsilon^\prime , S\cdot\tilde{\epsilon} ]
                      \left[ J_{2}^\prime ( \omega^\prime x , \tilde{m}^2 ) -
                       J_{2}^\prime ( - \omega^\prime x , \tilde{m}^2 ) \right]
                      \nonumber \\
                  & & \phantom{i \frac{g_{A}^2}{F_{\pi}^2} \int_{0}^{1}dx \;
                      \bar{u}_2(r^\prime) }
                      -\frac{1}{2}x\left(1-2x\right) \tilde{\epsilon} \cdot q
\;
                      \epsilon^\prime \cdot q \left[ J_{0}^\prime (
                      \omega^\prime x , \tilde{m}^2 ) +
J_{0}^\prime(-\omega^\prime x,
                      \tilde{m}^2 ) \right] \nonumber \\
                  & & \phantom{i \frac{g_{A}^2}{F_{\pi}^2} \int_{0}^{1}dx \;
                      \bar{u}_2(r^\prime) }
                      \left. + x\left(1-2x\right) [
S\cdot\epsilon^\prime,S\cdot q ]
                      \; \tilde{\epsilon} \cdot q \left[ J_{0}^\prime (
\omega^\prime x,
                      \tilde{m}^2 )
                      - J_{0}^\prime ( - \omega^\prime x ,
\tilde{m}^2)\right]\right\}
                      u_1(r) \nonumber\\
Amp_{7+8}^{N\pi}  &=& i \frac{g_{A}^2}{F_{\pi}^2} \int_{0}^{1}dx \int_{0}^{1}dy
                      \left(1-y\right)
                      \bar{u}_2(r^\prime) \times \nonumber \\
                  & & \phantom{i \frac{g_{A}^2}{F_{\pi}^2} }
                      \left\{ \tilde{\epsilon} \cdot \epsilon^\prime
                      \left[-2\left(d+1\right)\left(J_{6}^{\prime\prime}(T,\hat{m}^2)
                      + J_{6}^{\prime\prime} (-T,\hat{m}^2
)\right)\right.\right.
                      \nonumber \\
                  & & \phantom{i \frac{g_{A}^2}{F_{\pi}^2} \{ \tilde{\epsilon}
\cdot
                      \epsilon^\prime }
                      \left. + 2\left(T^2-(\hat{m}^2-m_{\pi}^2+\frac{q\cdot
q^\prime}{
                      \omega^\prime}T)\right)\left(J_{2}^{\prime\prime}(T,\hat{m}^2)+
                      J_{2}^{\prime\prime}(-T,\hat{m}^2)\right) \right]
                      \nonumber \\
                  & & \phantom{i \frac{g_{A}^2}{F_{\pi}^2} }
                      +\left[\left(1-d\right)\epsilon^\prime\cdot a\;
                      \tilde{\epsilon}\cdot b
                      -\epsilon^\prime\cdot\left(c+d\right)
\tilde{\epsilon}\cdot b-
                      2\;\epsilon^\prime\cdot a\;
                      \tilde{\epsilon}\cdot\left(c+d\right)\right]
                      \left( J_{2}^{\prime\prime} (T,
                      \hat{m}^2)+J_{2}^{\prime\prime}(-T,\hat{m}^2)\right)\nonumber\\
                  & & \phantom{i \frac{g_{A}^2}{F_{\pi}^2} }
                      +\left(T^2-(\hat{m}^2-m_{\pi}^2+\frac{q\cdot q^\prime}{
                      \omega^\prime} T)\right) \epsilon^\prime\cdot a\;
                      \tilde{\epsilon}\cdot b
                      \left(J_{0}^{\prime\prime} (T,\hat{m}^2)+
                      J_{0}^{\prime\prime}(-T,\hat{m}^2)\right) \nonumber \\
                  & & \phantom{i \frac{g_{A}^2}{F_{\pi}^2} }
                      +2\;\tilde{\epsilon}\cdot b\;
                      [S\cdot\epsilon^\prime,S\cdot(q-q^\prime)]
                      \left(J_{2}^{\prime\prime}(T,\hat{m}^2)-J_{2}^{\prime\prime}(
                      -T,\hat{m}^2)\right) \nonumber \\
                  & & \phantom{i \frac{g_{A}^2}{F_{\pi}^2} }
                      +4\;\epsilon^\prime\cdot a\;
                      [S\cdot\tilde{\epsilon},S\cdot(q-q^\prime)]
                      \left(J_{2}^{\prime\prime}(T,\hat{m}^2)-J_{2}^{\prime\prime}(
                      -T,\hat{m}^2)\right) \nonumber \\
                  & & \phantom{i \frac{g_{A}^2}{F_{\pi}^2} }
                      +4\;\tilde{\epsilon}\cdot\epsilon^\prime\;[S\cdot
c,S\cdot d]\left(
                      J_{2}^{\prime\prime} (T,\hat{m}^2)- J_{2}^{\prime\prime}(
                      -T,\hat{m}^2)\right) \nonumber \\
                  & & \phantom{i \frac{g_{A}^2}{F_{\pi}^2} }\left.
                      +2\; \epsilon^\prime\cdot a\;\tilde{\epsilon}\cdot b\;
                      [S\cdot c,S\cdot d]
                      \left(J_{0}^{\prime\prime}(T,\hat{m}^2)-J_{0}^{\prime\prime}(
                      -T,\hat{m}^2)\right) \right\} u_1(r) \nonumber\\
Amp_{9}^{N\pi}    &=& i \frac{g_{A}^2}{F_{\pi}^2} \; \bar{u}_2(r^\prime) \;
                      \tilde{\epsilon} \cdot \epsilon^\prime \; u_1(r)
\int_{0}^{1}dx
                      \left\{ \left(d-1\right) J_{2}^\prime ( 0 , m_{f}^2 ) +
                      \left(m_{f}^2-m_{\pi}^2\right) J_{0}^\prime ( 0 ,m_{f}^2)
                      \right\} ,
\end{eqnarray}
with
\begin{eqnarray}
a_\mu&=&-q_\mu \; y \nonumber\\
b_\mu&=&q^{\prime}_\mu\left(2y+2x-2y x-2\right)-q_\mu\left(2y-1\right)
\nonumber\\
c_\mu&=&q^{\prime}_\mu\left(y+x-y x\right)-q_\mu y \nonumber\\
d_\mu&=&q^{\prime}_\mu\left(y+x-y x-1\right)-q_\mu\left(y-1\right),
\end{eqnarray}
and the energy and mass variables $T,\hat{m},\tilde{m},m_f$ as defined in
section \ref{chiamp}.

\section{$\Delta\pi$ Loop Amplitudes in VCS}\label{appdpi}

The 9 $\pi\Delta$ continuum diagrams are shown in Fig.\ref{figgdpi}. We find
\begin{eqnarray}
Amp_{1+2}^{\Delta\pi}&=& i \frac{8 g_{\pi N\Delta}^2}{3 F_{\pi}^2} \;
\bar{u}_2(r^\prime)
     \left\{ -{1\over 2}\frac{d-2}{d-1} \tilde{\epsilon} \cdot \epsilon^\prime
            \left[J_0(\omega^\prime -\Delta ,m_{\pi}^2)+J_0(-\omega^\prime
            -\Delta, m_{\pi}^2)\right]\right.\nonumber\\
&-&\left.{1\over 2}\frac{2}{d-1}\;[S\cdot\epsilon^\prime,
                      S\cdot\tilde{\epsilon}]\left[J_0(\omega^\prime -\Delta
,m_{\pi}^2)-
                      J_0(-\omega^\prime -\Delta,m_{\pi}^2)\right] \right\}
u_1(r)  \nonumber\\
Amp_{3+6}^{\Delta\pi}  &=& i {8g_{\pi N\Delta}^2\over 3F_\pi^2} \int_{0}^{1}dx
\;
                      \bar{u}_2(r^\prime) \left\{ {d-2\over d-1}
                      \tilde{\epsilon} \cdot\epsilon^\prime
                      \left[ J^{\prime}_2 ( \omega^\prime x -\Delta, m_{\pi}^2
) +
                      J^{\prime}_2 ( -\omega^\prime x-\Delta, m_{\pi}^2 )
                      \right] \right. \nonumber \\
                  & & \phantom{i \frac{g_{A}^2}{F_{\pi}^2} \int_{0}^{1}dx \;
                      \bar{u}_2(r^\prime) } \left.+{2\over d-1} \;
                   [S\cdot\epsilon^\prime,
                      S\cdot \tilde{\epsilon} ] \left[ J^{\prime}_2 (
\omega^\prime
                   x-\Delta,m_{\pi}^2
                      ) - J^{\prime}_2 ( -\omega^\prime x-\Delta, m_{\pi}^2 )
                      \right] \right\} u_1(r)  \nonumber\\
Amp_{4+5}^{\Delta\pi}  &=& i {8g_{\pi N\Delta}^2\over 3F_\pi^2} \int_{0}^{1}dx
\;
                      \bar{u}_2(r^\prime) \left\{{d-2\over d-1}
                      \tilde{\epsilon} \cdot \epsilon^\prime
                      \left[ J_{2}^\prime ( \omega^\prime x -\Delta,
\tilde{m}^2 ) +
                       J_{2}^\prime ( - \omega^\prime x -\Delta,\tilde{m}^2
                      )\right]\right.
                      \nonumber \\
                  & & \phantom{i \frac{g_{A}^2}{F_{\pi}^2} \int_{0}^{1}dx \;
                      \bar{u}_2(r^\prime) }
                      +{2\over d-1} \;
                      [ S\cdot \epsilon^\prime , S\cdot\tilde{\epsilon} ]
                      \left[ J_{2}^\prime ( \omega^\prime x-\Delta ,
\tilde{m}^2 ) -
                       J_{2}^\prime ( - \omega^\prime x-\Delta , \tilde{m}^2 )
\right]
                      \nonumber \\
                  & & \phantom{i \frac{g_{A}^2}{F_{\pi}^2} \int_{0}^{1}dx \;
                      \bar{u}_2(r^\prime) }
                      -\frac{1}{2}{d-2\over d-1}
                      x\left(1-2x\right) \tilde{\epsilon} \cdot q^\prime \;
                      \epsilon^\prime \cdot q \left[ J_{0}^\prime (
                      \omega^\prime x -\Delta, \tilde{m}^2 ) + J_{0}^\prime
                      (-\omega^\prime x-\Delta,
                      \tilde{m}^2 ) \right] \nonumber \\
                  & & \phantom{i \frac{g_{A}^2}{F_{\pi}^2} \int_{0}^{1}dx \;
                      \bar{u}_2(r^\prime) }
                      \left. -{1\over d-1} x\left(1-2x\right) [
S\cdot\epsilon^\prime
                      ,S\cdot q ]
                      \; \tilde{\epsilon} \cdot q \left[ J_{0}^\prime (
\omega^\prime
                      x-\Delta,
                      \tilde{m}^2 )
                      - J_{0}^\prime ( - \omega^\prime x -\Delta, \tilde{m}^2
                      )\right]\right\}
                      u_1(r) \nonumber\\
Amp_{7+8}^{\Delta\pi}  &=& i {8g_{\pi N\Delta}^2\over 3F_\pi^2}
                      \int_{0}^{1}dx \int_{0}^{1}dy
                      \left(1-y\right)
                      \bar{u}_2(r^\prime) \times \nonumber \\
                  & & \phantom{i \frac{g_{A}^2}{F_{\pi}^2} }
                      \left\{ \tilde{\epsilon} \cdot \epsilon^\prime
                      \left[-2{(d+1)(d-2)\over d-1}
                      \left(J_{6}^{\prime\prime}(T-\Delta,\hat{m}^2)
                      + J_{6}^{\prime\prime} (-T-\Delta,\hat{m}^2 )
                      \right)\right.\right.
                      \nonumber \\
                  & & \phantom{i \frac{g_{A}^2}{F_{\pi}^2} \{ \tilde{\epsilon}
\cdot
                      \epsilon^\prime }
                      \left. + 2{d-2\over
d-1}\left(T^2-(\hat{m}^2-m_{\pi}^2+\frac{q\cdot
                      q^\prime}{
                      \omega^\prime}
                       T)\right)\left(J_{2}^{\prime\prime}(T-\Delta,\hat{m}^2)+
                      J_{2}^{\prime\prime}(-T-\Delta,\hat{m}^2)\right) \right]
                      \nonumber \\
                  & & \phantom{i \frac{g_{A}^2}{F_{\pi}^2} }
                      +{d-2\over d-1}\left[\left(1-d\right)\epsilon^\prime\cdot
                       a\;\tilde{\epsilon}\cdot b
                      -\epsilon^\prime\cdot\left(c+d\right)
\tilde{\epsilon}\cdot b-
                      2\;\epsilon^\prime\cdot
                       a\;\tilde{\epsilon}\cdot\left(c+d\right)\right]
                      \left( J_{2}^{\prime\prime} (T-\Delta,
                      \hat{m}^2)+J_{2}^{\prime\prime}
                      (-T-\Delta,\hat{m}^2)\right)\nonumber\\
                  & & \phantom{i \frac{g_{A}^2}{F_{\pi}^2} }
                      +{d-2\over d-1}
                      \left(T^2-(\hat{m}^2-m_{\pi}^2+\frac{q\cdot q^\prime}{
                      \omega^\prime} T)\right) \epsilon^\prime\cdot
                       a\;\tilde{\epsilon}\cdot b
                      \left(J_{0}^{\prime\prime} (T-\Delta,\hat{m}^2)+
                      J_{0}^{\prime\prime}(-T-\Delta,\hat{m}^2)\right)
\nonumber \\
                  & & \phantom{i \frac{g_{A}^2}{F_{\pi}^2} }
                      -{2\over d-1}\;\tilde{\epsilon}\cdot
                      b\;[S\cdot\epsilon^\prime,S\cdot(q-q^\prime)]
                      \left(
                      J_{2}^{\prime\prime}(T-\Delta,\hat{m}^2)-J_{2}^{\prime\prime}(
                      -T-\Delta,\hat{m}^2)\right) \nonumber \\
                  & & \phantom{i \frac{g_{A}^2}{F_{\pi}^2} }
                      -{4\over d-1}\;\epsilon^\prime\cdot
                       a\;[S\cdot\tilde{\epsilon},S\cdot(q-q^\prime)]
                      \left(
                       J_{2}^{\prime\prime}(T-\Delta,\hat{m}^2)-J_{2}^{\prime\prime}(
                      -T-\Delta,\hat{m}^2)\right) \nonumber \\
                  & & \phantom{i \frac{g_{A}^2}{F_{\pi}^2} }
                      -{4\over d-1}\;\tilde{\epsilon}\cdot\epsilon^\prime\;
                       [S\cdot c,S\cdot d]\left(
                      J_{2}^{\prime\prime} (T-\Delta,\hat{m}^2)-
J_{2}^{\prime\prime}(
                      -T-\Delta,\hat{m}^2)\right) \nonumber \\
                  & & \phantom{i \frac{g_{A}^2}{F_{\pi}^2} }\left.
                      -{2\over d-1}\; \epsilon^\prime\cdot a\;
                       \tilde{\epsilon}\cdot b\;[S\cdot
                       c,S\cdot d]
                      \left(J_{0}^{\prime\prime}(T-\Delta,\hat{m}^2)
                      -J_{0}^{\prime\prime}(
                      -T-\Delta,\hat{m}^2)\right) \right\} u_1(r) \nonumber\\
Amp_{9}^{\Delta\pi}    &=& i {8g_{\pi N\Delta}^2\over 3F_\pi^2}\;
\bar{u}_2(r^\prime) \;
                      \tilde{\epsilon} \cdot \epsilon^\prime \; u_1(r)
\int_{0}^{1}dx
                      \left\{ \left(d-2\right) J_{2}^\prime ( -\Delta , m_{f}^2
) +
                      {d-2\over d-1}\left(m_{f}^2-m_{\pi}^2\right) J_{0}^\prime
                      ( -\Delta ,m_{f}^2)
                      \right\} ,
\end{eqnarray}

\section{$\pi^0$-Pole Contributions}\label{anomalypolas}

In this section we explicitly give the ${\cal O}(p^3)\equiv {\cal
O}(\epsilon^3)$
contribution of $\pi^0$-exchange in the t-channel---Fig.\ref{figgborn}f---to
the generalized spin-polarizabilities of
Eqs.(\ref{4spin},\ref{3spin}). In the main part of this work we had included
this
particular effect in the Born part of the structure amplitudes
$A_i^{Born}(\omega^\prime,\theta,\bar{q})$ (c.f. Eq.(\ref{eq:bg})). However,
in the existing literature of VCS many authors prefer to
consider
$\pi^0$-exchange as a genuine contribution to the spin-polarizabilities. For
easier
comparison we list our results below and show the resulting
GPs in Fig.\ref{figanom1}.
\begin{eqnarray}
\alpha_E(\bar{q})&=&\beta_M(\bar{q})=0\nonumber\\
\hat{P}^{\rm anom}_{(01,1)1}&=&\sqrt{2\over 3}{g_A\over 24\pi^2F_\pi^2}
{1\over m_\pi^2+\bar{q}^2}\nonumber\\
P_{(01,12)1}^{\rm anom}&=&-\sqrt{2}{g_A\over 24\pi^2F_\pi^2}
{1\over m_\pi^2+\bar{q}^2}\nonumber\\
P_{(11,02)1}^{\rm anom}&=&\sqrt{2\over 3}{g_A\over 12\pi^2F_\pi^2}
{1\over m_\pi^2+\bar{q}^2}\nonumber\\
P_{(11,00)1}^{\rm anom}&=&-\sqrt{1\over 3}{g_A\over 12\pi^2F_\pi^2}
{\bar{q}^2\over m_\pi^2+\bar{q}^2}\nonumber\\
\hat{P}^{\rm anom}_{(11,2)1}&=&\hat{P}^{\rm anom}_{(01,1)0}=
P^{\rm anom}_{(01,01)1}=0\nonumber\\
P^{\rm anom}_{(11,11)1}&=&-\omega_0{g_A\over 12\pi^2F_\pi^2}{1\over
m_\pi^2+\bar{q}^2}
\end{eqnarray}

\section{A useful identity}\label{ident}

It should be noted 
that, while making the transition from the chiral loop amplitudes in appendices
\ref{appnpi}, \ref{appdpi} 
to the twelve VCS structure amplitudes $A_i,\;i=1\dots 12$ of 
Eq.(\ref{eq:vcs12}), 
one also encounters the matrix element
\begin{equation}
\vec{\epsilon}^{\;\prime}\cdot\hat{q}\;\vec{\epsilon}\cdot\hat{q}^\prime\;\vec{\sigma}
\cdot\left(\hat{q}^\prime\times\hat{q}\right)=-\left(\vec{\epsilon}^{\;\prime}\times
\vec{\epsilon}\right)\cdot\left(\hat{q}^\prime\times\hat{q}\right)\;\vec{\sigma}\cdot
\left(\hat{q}^\prime\times\hat{q}\right),
\label{eq:cross}
\end{equation}
which has to be brought into a form which accompanies one of the twelve
structure
amplitudes. To achieve this we start from the identity
\begin{eqnarray}
\vec{a}\cdot\hat{e}_x\;\vec{b}\cdot\hat{e}_x+\vec{a}\cdot\hat{e}_y\;\vec{b}\cdot
\hat{e}_y+\vec{a}\cdot\hat{e}_z\;\vec{b}\cdot\hat{e}_z=\vec{a}\cdot\vec{b}
\nonumber
\end{eqnarray}
and then construct the 3 orthonormal unit vectors $\hat{e}_a ,\;a=x,y,z$ from
the direction vectors $\hat{q},\hat{q}^\prime$ via
\begin{eqnarray}
\hat{e}_x=\frac{1}{\sin\theta}\left(\hat{q}^\prime\times\hat{q}\right)\times\hat{q}
\; ,\quad
\hat{e}_y=\frac{1}{\sin\theta}\left(\hat{q}^\prime\times\hat{q}\right) , \quad
\hat{e}_z=\hat{q} . \nonumber
\end{eqnarray}
Identifying
$\vec{a}=(\vec{\epsilon}^{\;\prime}\times\vec{\epsilon})\,{\rm and}\,
\vec{b}=\vec{\sigma}$
one finds a relation for the structure of interest, Eq.(\ref{eq:cross}),
\begin{eqnarray}
\left(\vec{\epsilon}^{\;\prime}\times\vec{\epsilon}\right)\cdot\left(\hat{q}^\prime\times
\hat{q}\right)\;\vec{\sigma}\cdot\left(\hat{q}^\prime\times\hat{q}\right)=
\sin^2\theta\;\sigma\cdot\left(\vec{\epsilon}^{\;\prime}\times\vec{\epsilon}\right)-
\vec{a}\cdot\vec{c}\;\vec{b}\cdot\vec{c}-\sin^2\theta\left(\vec{\epsilon}^{\;\prime}
\times\vec{\epsilon}\right)\cdot\hat{q}\;\vec{\sigma}\cdot\hat{q} \; ,
\nonumber
\end{eqnarray}
with $\vec{c}=\hat{q}\;\cos\theta-\hat{q}^\prime$. Noting that
\begin{eqnarray}
\vec{a}\cdot\vec{c}\;\vec{b}\cdot\vec{c}&=&\sin^2\theta\;\vec{a}\cdot\vec{b}+\left(
\vec{a}\times\vec{c}\right)\cdot\left(\vec{c}\times\vec{b}\right), \nonumber \\
\left(\vec{\epsilon}^{\;\prime}\times\vec{\epsilon}\right)\cdot\hat{q}\;\vec{\sigma}\cdot
\hat{q}&=&-\vec{\epsilon}\cdot\hat{q}\;\vec{\sigma}\cdot\left(\vec{\epsilon}^{\;\prime}
\times\hat{q}\right)+\vec{\epsilon}^{\;\prime}\cdot\hat{q}\;\vec{\sigma}\cdot\left(
\vec{\epsilon}\times\hat{q}\right)+\vec{\sigma}\cdot\left(\vec{\epsilon}^{\;\prime}\times
\vec{\epsilon}\right), \nonumber
\end{eqnarray}
one obtains
\begin{eqnarray}
\left(\vec{\epsilon}^{\;\prime}\times\vec{\epsilon}\right)\cdot\left(\hat{q}^\prime\times
\hat{q}\right)\;\vec{\sigma}\cdot\left(\hat{q}^\prime\times\hat{q}\right)&=&
\vec{\epsilon}\cdot\hat{q}\;\vec{\sigma}\cdot\left(\vec{\epsilon}^{\;\prime}\times\hat{q}
\right)-\vec{\epsilon}^{\;\prime}\cdot\hat{q}\;\vec{\sigma}\cdot\left(\vec{\epsilon}
\times\hat{q}\right)-\cos\theta\;\vec{\epsilon}\cdot\hat{q}\;\vec{\sigma}\cdot\left(\vec{
\epsilon}^\prime\times\hat{q}^\prime\right)-\cos\theta\;\vec{\epsilon}\cdot
\hat{q}^\prime\;\vec{\sigma}\cdot\left(\vec{\epsilon}^{\;\prime}\times\hat{q}\right)
\nonumber \\
& &+\;\vec{\epsilon}\cdot\hat{q}^\prime\;\vec{\sigma}\cdot\left(\vec{\epsilon}^{\;\prime}
\times\hat{q}^\prime\right)+\cos\theta\;\vec{\epsilon}^{\;\prime}\cdot\hat{q}\;\vec{
\sigma}\cdot\left(\vec{\epsilon}\times\hat{q}^\prime\right)-\sin^2\theta\;\vec{\sigma}
\cdot\left(\vec{\epsilon}^{\;\prime}\times\vec{\epsilon}\right) .
\end{eqnarray}

\end{appendix}

\bigskip\bigskip
\newpage



\newpage
\begin{table}
\begin{center}
\begin{tabular}{ccccccc}
\hline
Quantity& Expt.& ChPT&LSM&ELM&NRQM\\
$P_{LL}-{1\over \epsilon}P_{TT}$&$30.5\pm 6.2$
&26.3&10.9&5.9&17.0\\
$P_{LT}$&$-8.6\pm 3.9$&-5.7&0&-1.9&-1.7\\
\hline
\end{tabular}
\vspace{0.5cm}
\caption{Experimental values of the response functions measured at MAMI
at $Q^2=0.33$ GeV$^2$ compared with
predictions from chiral perturbation theory at ${\cal O}(p^3)$,
the linear sigma model (LSM) of Metz and Drechsel \protect 
\cite{Metz}, 
the effective lagrangian model (ELM) of Vanderhaeghen \protect \cite{Vanderhaeghen},
and the nonrelativistic quark model (NRQM) of Guichon et al. \protect \cite{Guichon}.
This table is taken from ref.\protect \cite{nstar}.
}
\end{center}
\end{table}
\newpage

\begin{figure}[t]
\centerline{\epsfig{file=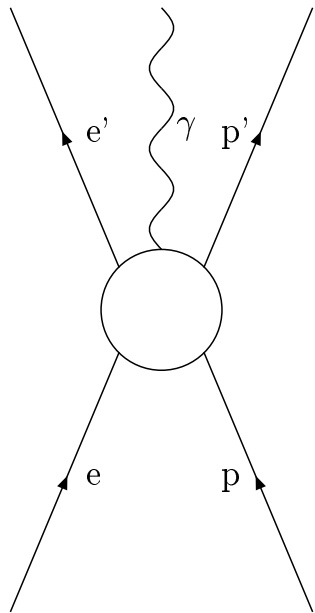,height=8cm}}
\vspace{5cm}
\caption[diag]{\label{process} The process $e\,p\rightarrow e^\prime
p^\prime\gamma$.}
\end{figure}
\newpage

\begin{figure}[t]
\centerline{\epsfig{file=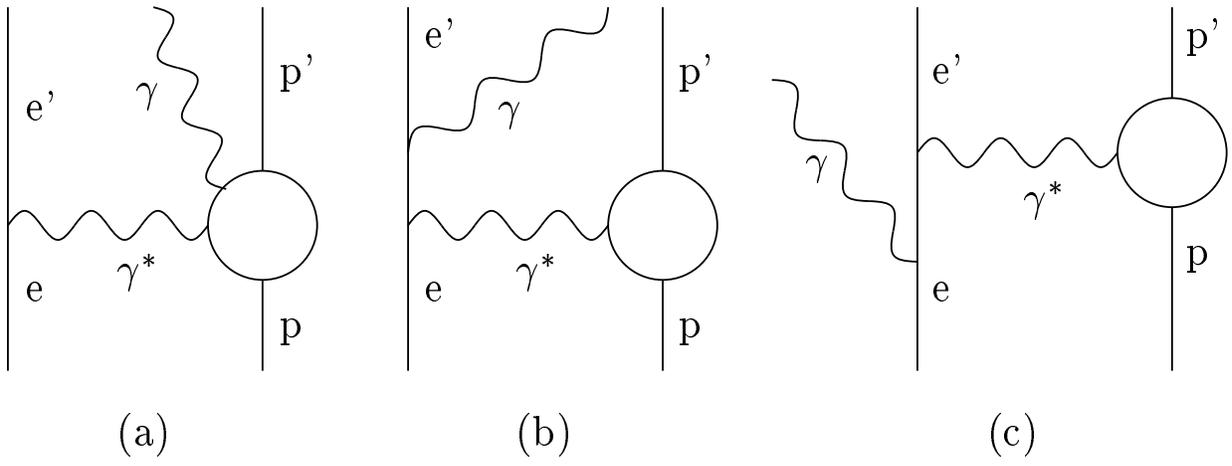,height=6cm}}
\vspace{3cm}
\caption[diag]{\label{vcsbh}
Genuine diagrams for the proper VCS process (a) and for
the associated Bethe-Heitler corrections (b,c).}
\end{figure}
\newpage

\begin{figure}[t]
\centerline{\psfig{file=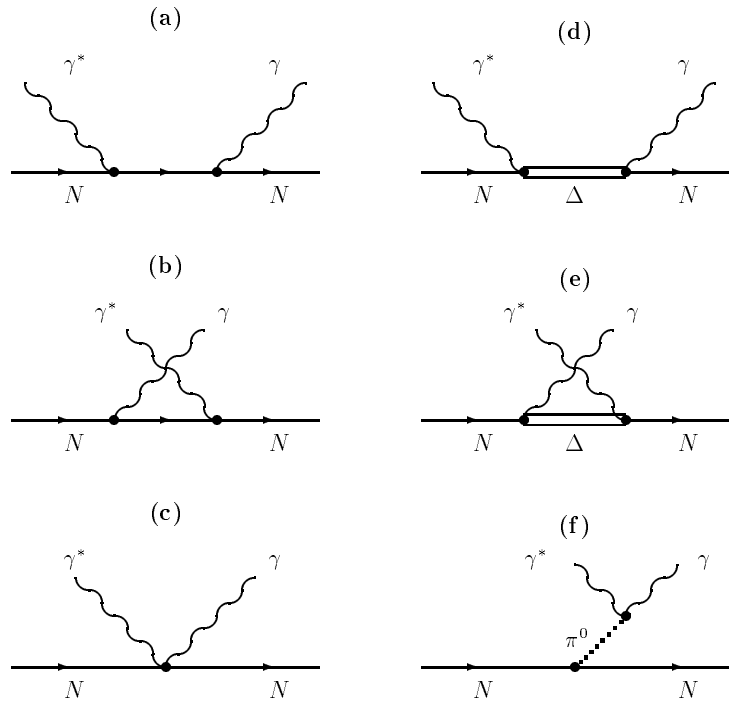,height=23cm}}
\vspace{-6cm}
\caption[diag]{\label{figgborn} Born diagrams for VCS in the ``Small Scale
Expansion''.}
\end{figure}
\newpage

\begin{figure}[t]
\centerline{\psfig{file=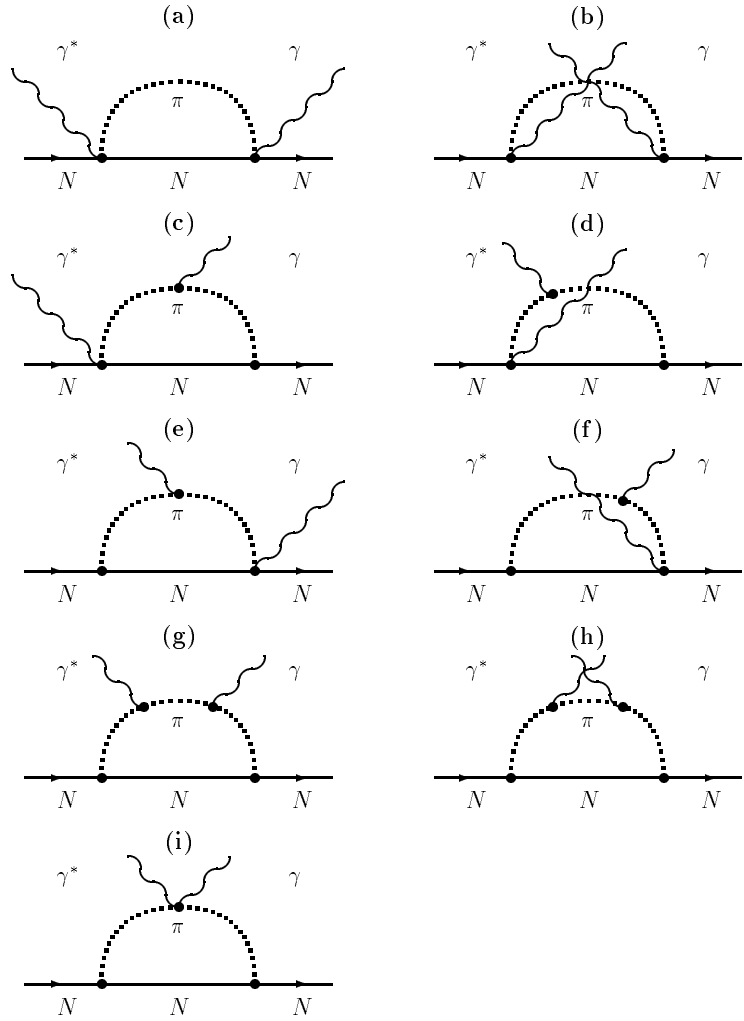,height=23cm}}
\vspace{-5cm}
\caption[diag]{\label{figgnpi} ${\cal O}(p^3)$ $N\pi$-loop diagrams for VCS.}
\end{figure}
\newpage

\begin{figure}[t]
\centerline{\psfig{file=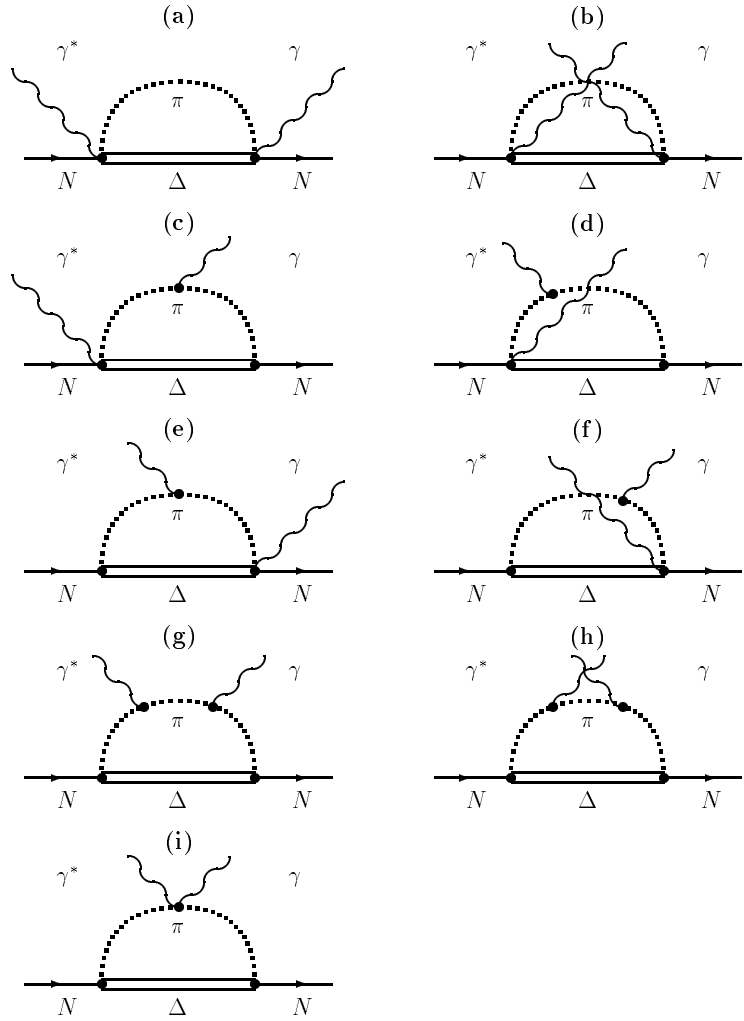,height=23cm}}
\vspace{-5cm}
\caption[diag]{\label{figgdpi} ${\cal O}(\epsilon^3)$ $\Delta\pi$-loop diagrams
for VCS.}
\end{figure}
\newpage

\begin{figure}[h]
\psfrag{xx1}{$\bar{q}^2\; [{\rm GeV}^2]$}
\psfrag{xx2}{$\bar{q}^2\; [{\rm GeV}^2]$}
\psfrag{yy1}{$\bar{\alpha}_E^{(3)} (\bar{q}^2)\quad [10^{-4}\,{\rm fm}^3]$}
\psfrag{yy2}{$\bar{\beta}_M^{(3)} (\bar{q}^2)\quad [10^{-4}\,{\rm fm}^3]$}
\centerline{\epsfig{file=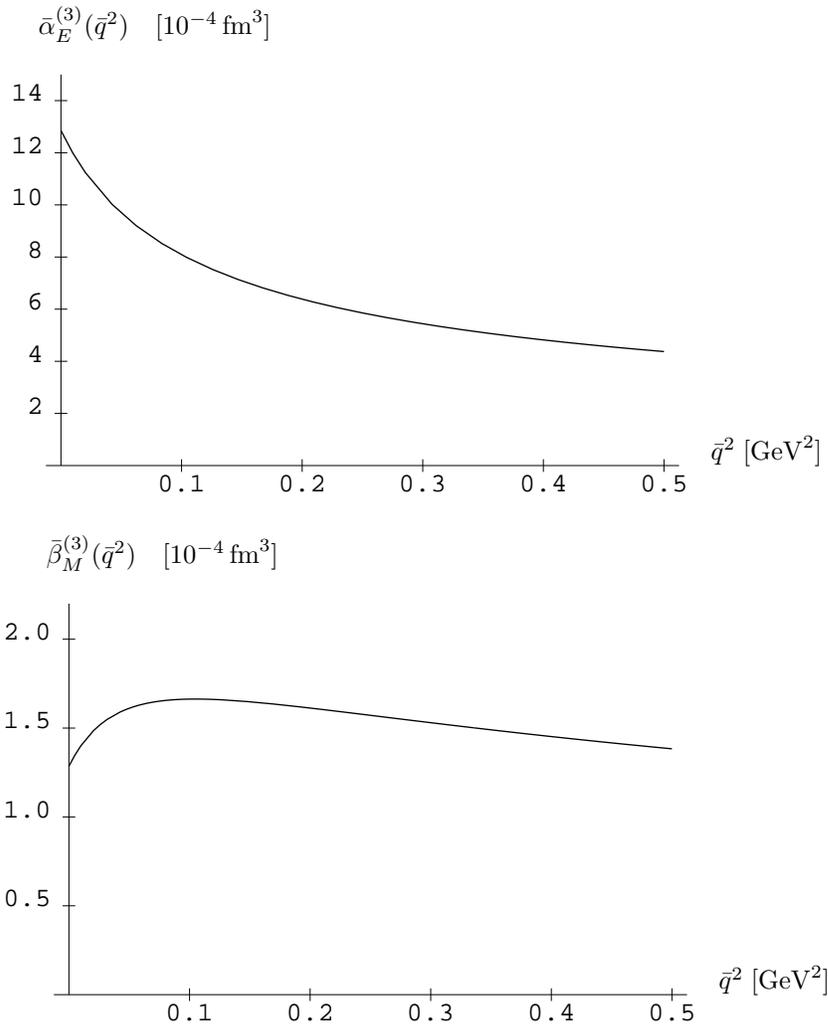}}
\vspace{2.5cm}
\caption[diag]{\label{figab} ${\cal O}(p^3)$ HBChPT results for the two
spin-independent generalized polarizabilities $\bar{\alpha}_E(\bar{q}^2),\;
\bar{\beta}_M(\bar{q}^2)$ of Eq.(\ref{eq:bq}).}
\end{figure}

\newpage

\begin{figure}[h]
\psfrag{xx3}{$\bar{q}^2\; [{\rm GeV}^2]$}
\psfrag{yy343}{$P^{(3)}_{\left(01,12\right)1} (\bar{q}^2)\quad [10^{-3}\,{\rm
fm}^4]$}
\psfrag{yy344}{$P^{(3)}_{\left(11,02\right)1} (\bar{q}^2)\quad [10^{-3}\,{\rm
fm}^4]$}
\psfrag{yy225}{$P^{(3)}_{\left(11,00\right)1} (\bar{q}^2)\quad [10^{-2}\,{\rm
fm}^2]$}
\psfrag{yy346}{$\hat{P}^{(3)}_{\left(01,1\right)1}(\bar{q}^2)\quad
[10^{-3}\,{\rm fm}^4]$}
\vspace{2cm}
\centerline{\epsfig{file=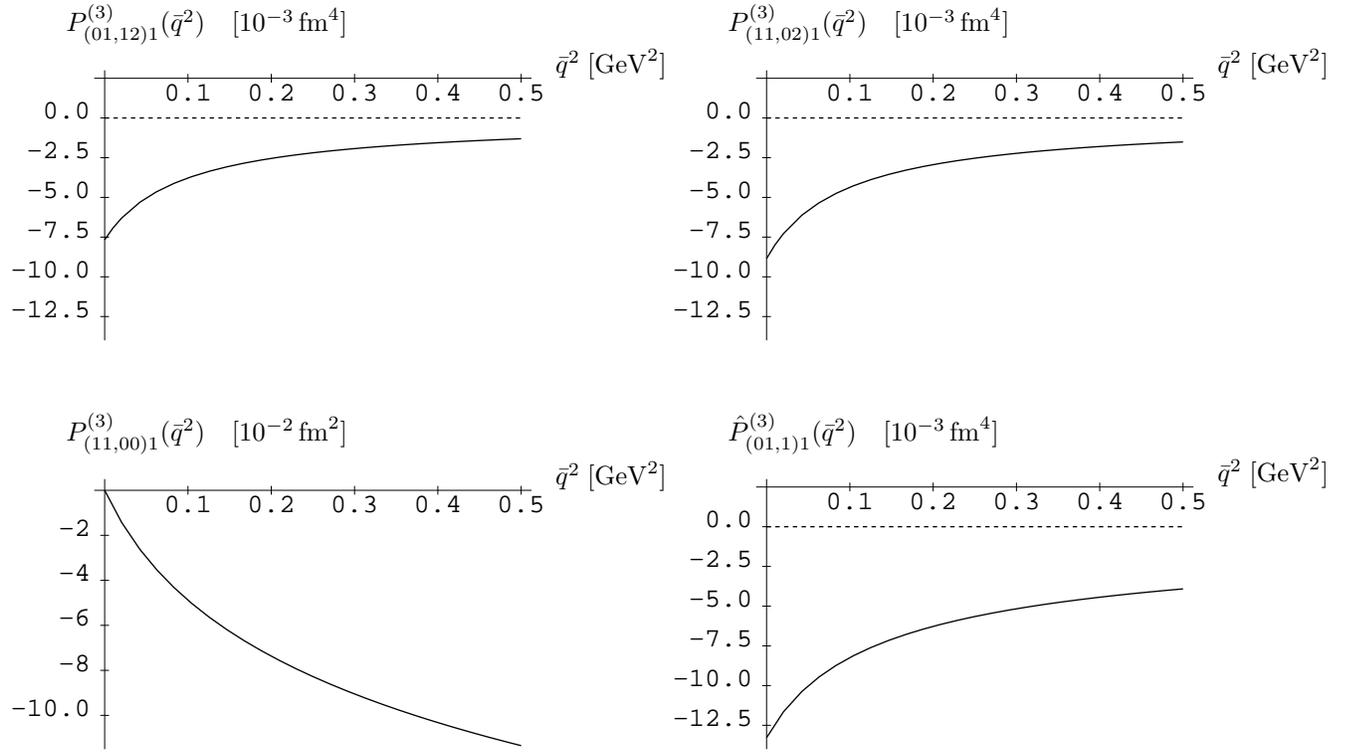}}
\vspace{1.5cm}
\caption[diag]{\label{figspin1} ${\cal O}(p^3)$ HBChPT results for
the four
independent generalized spin polarizabilities of Eq.(\ref{4spin}).
Note that the ``anomaly-contributions'' of appendix \ref{anomalypolas} 
are not included here but plotted separately
in Fig.\ref{figanom1}.}
\end{figure}

\newpage

\begin{figure}[h]
\psfrag{xx3}{$\bar{q}^2\; [{\rm GeV}^2]$}
\psfrag{yy33a}{$P^{l.o.\chi}_{\left(01,01\right)1} (\bar{q}^2)\quad [10^{-3}\,{\rm
fm}^3]$}
\psfrag{yy33b}{$P^{l.o.\chi}_{\left(11,11\right)1} (\bar{q}^2)\quad [10^{-3}\,{\rm
fm}^3]$}
\psfrag{yy45}{$\hat{P}^{l.o.\chi}_{\left(11,2\right)1} (\bar{q}^2)\quad [10^{-4}\,
{\rm fm}^5]$}
\vspace{2cm}
\centerline{\epsfig{file=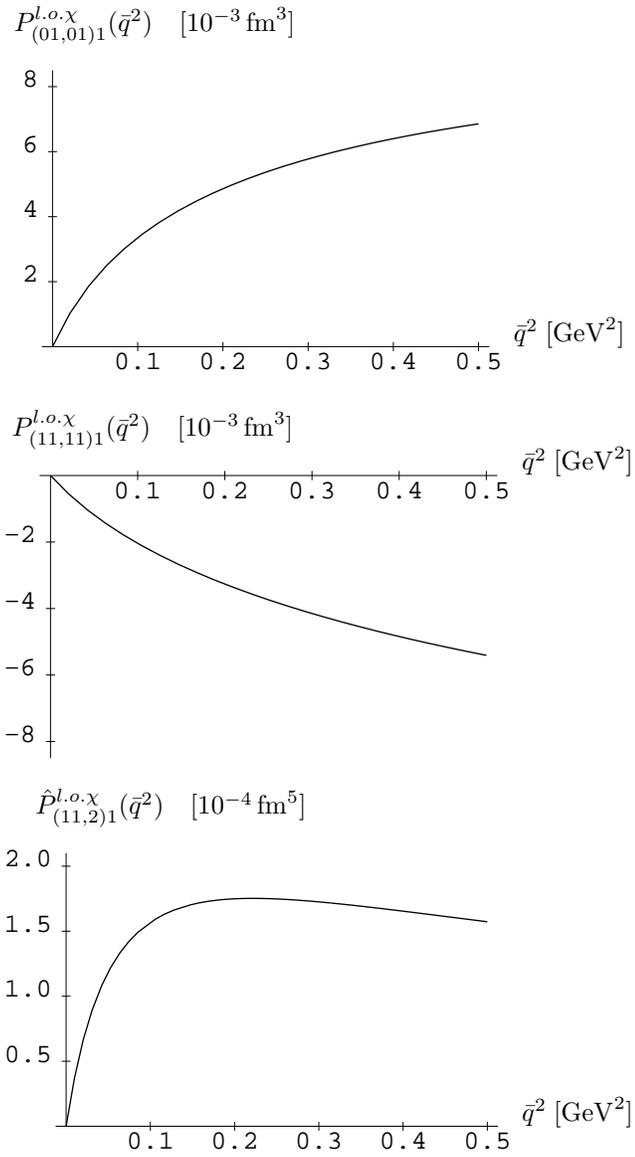}}
\vspace{2.5cm} 
\caption[diag]{\label{figspin2} ${\cal O}(p^3)$ HBChPT results for the three
redundant spin polarizabilities of Eq.(\ref{3spin}), reconstructed from the
C-invariance constraints of Eq.(\ref{cinv}). Note that the ``anomaly-contributions'' 
of appendix \ref{anomalypolas} are not included here but plotted separately
in Fig.\ref{figanom1}.}
\end{figure}
\newpage

\begin{figure}[h]
\psfrag{xx1}{$\bar{q}^2\; [{\rm GeV}^2]$}
\psfrag{xx2}{$\bar{q}^2\; [{\rm GeV}^2]$}
\psfrag{yy1}{$\bar{\alpha}_E^{(III)} (\bar{q}^2)\quad [10^{-4}\,{\rm fm}^3]$}
\psfrag{yy2}{$\bar{\beta}_M^{(III)} (\bar{q}^2)\quad [10^{-4}\,{\rm fm}^3]$}
\centerline{\epsfig{file=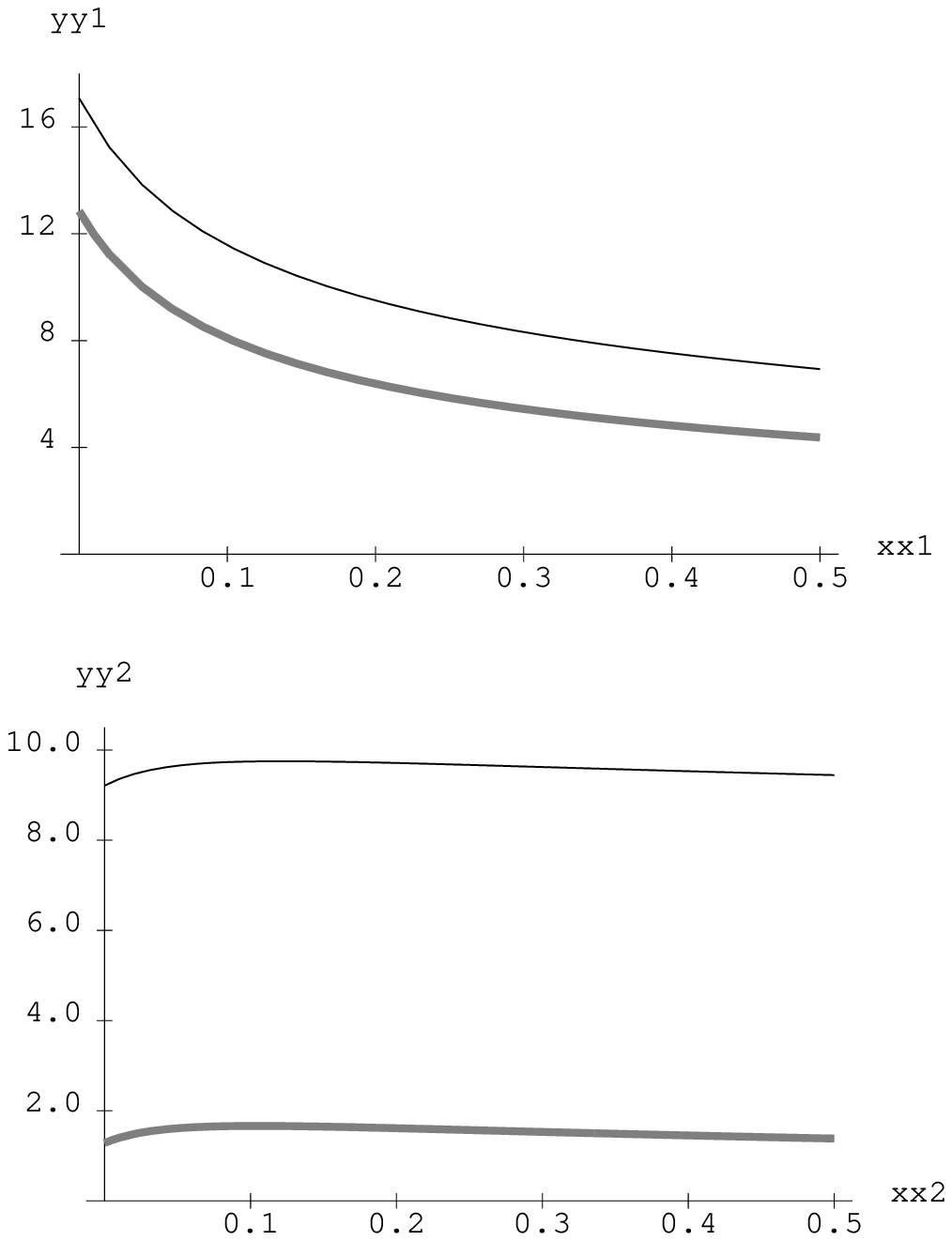}}
\vspace{4.5cm}
\caption[diag]{\label{figabSSE} Absolute ${\cal O}(\epsilon^3)$ SSE results for the two
spin-independent generalized polarizabilities $\bar{\alpha}_E^{(III)}(\bar{q}^2),\;
\bar{\beta}_M^{(III)}(\bar{q}^2)$ of Eq.(\ref{eq:bqSSE}), compared to
the ${\cal O}(p^3)$ HBChPT results shown in gray shading.}
\end{figure}
\newpage

\begin{figure}[h]
\psfrag{xx1}{$\bar{q}^2\; [{\rm GeV}^2]$}
\psfrag{xx2}{$\bar{q}^2\; [{\rm GeV}^2]$}
\psfrag{yy1}{$\bar{\alpha}_E^{ren.} (\bar{q}^2)\quad [10^{-4}\,{\rm fm}^3]$}
\psfrag{yy2}{$\bar{\beta}_M^{ren.} (\bar{q}^2)\quad [10^{-4}\,{\rm fm}^3]$}
\centerline{\epsfig{file=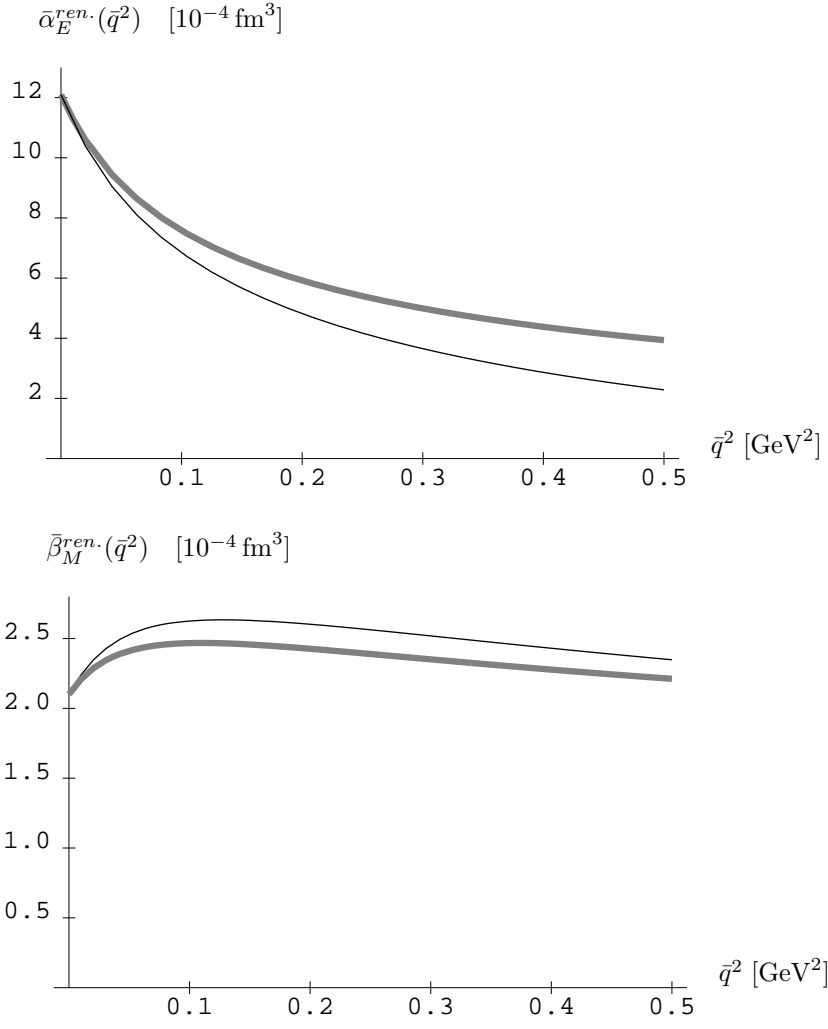}}
\vspace{4.5cm}
\caption[diag]{\label{figabren} Comparison between the ${\cal O}(\epsilon^3)$ SSE 
results for $\bar{\alpha}_E^{ren.}(\bar{q}^2),\;
\bar{\beta}_M^{ren.}(\bar{q}^2)$ with ${\cal O}(p^3)$ HBCHPT (in gray shading). 
Note that 
all curves have been normalized to the experimental results of 
$\bar{\alpha}_E,\;\bar{\beta}_M$ of the proton at the real photon point 
$\bar{q}\rightarrow 0$.}
\end{figure}

\newpage

\begin{figure}[h]
\psfrag{xx3}{$\bar{q}^2\; [{\rm GeV}^2]$}
\psfrag{yy343}{$P^{(III)}_{\left(01,12\right)1} (\bar{q}^2)\quad [10^{-3}\,{\rm
fm}^4]$}
\psfrag{yy344}{$P^{(III)}_{\left(11,02\right)1} (\bar{q}^2)\quad [10^{-3}\,{\rm
fm}^4]$}
\psfrag{yy225}{$P^{(III)}_{\left(11,00\right)1} (\bar{q}^2)\quad [10^{-2}\,{\rm
fm}^2]$}
\psfrag{yy346}{$\hat{P}^{(III)}_{\left(01,1\right)1}(\bar{q}^2)\quad
[10^{-3}\,{\rm fm}^4]$}
\vspace{2cm}
\centerline{\epsfig{file=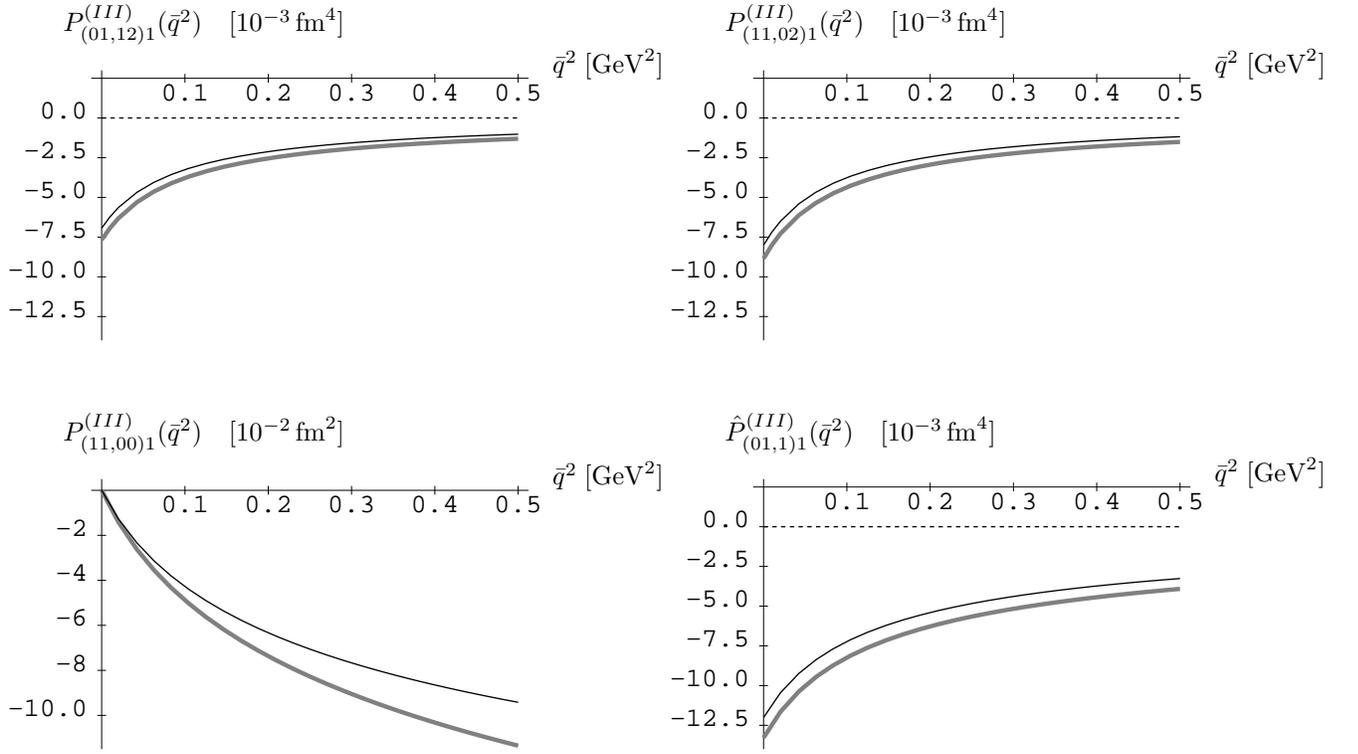}}
\vspace{1.5cm}
\caption[diag]{\label{figSSEspin1} ${\cal O}(\epsilon^3)$ SSE results for
the four
independent generalized spin polarizabilities of Eq.(\ref{sdspin}), 
compared to the ${\cal O}(p^3)$ HBChPT results of Eq.(\ref{4spin})
in gray shading. Note that the ``anomaly-contributions'' 
of appendix \ref{anomalypolas} are not included but plotted separately
in Fig.\ref{figanom1}.}
\end{figure}
\newpage


\begin{figure}[h]
\psfrag{xx3}{$\bar{q}^2\; [{\rm GeV}^2]$}
\psfrag{yy341}{$P^{anom.}_{\left(01,12\right)1} (\bar{q}^2)\quad [10^{-3}\,{\rm
fm}^4]$}
\psfrag{yy342}{$P^{anom.}_{\left(11,02\right)1} (\bar{q}^2)\quad [10^{-3}\,{\rm
fm}^4]$}
\psfrag{yy223}{$P^{anom.}_{\left(11,00\right)1} (\bar{q}^2)\quad [10^{-2}\,{\rm
fm}^2]$}
\psfrag{yy344}{$\hat{P}^{anom.}_{\left(01,1\right)1}(\bar{q}^2)\quad [10^{-3}\,
{\rm fm}^4]$}
\psfrag{yy335}{$P^{anom.}_{\left(01,01\right)1} (\bar{q}^2)\quad [10^{-3}\,{\rm
fm}^3]$}
\psfrag{yy336}{$P^{anom.}_{\left(11,11\right)1} (\bar{q}^2)\quad [10^{-3}\,{\rm
fm}^3]$}
\psfrag{yy457}{$\hat{P}^{anom.}_{\left(11,2\right)1} (\bar{q}^2)\quad
[10^{-4}\,
{\rm fm}^5]$}
\vspace{2cm}
\centerline{\epsfig{file=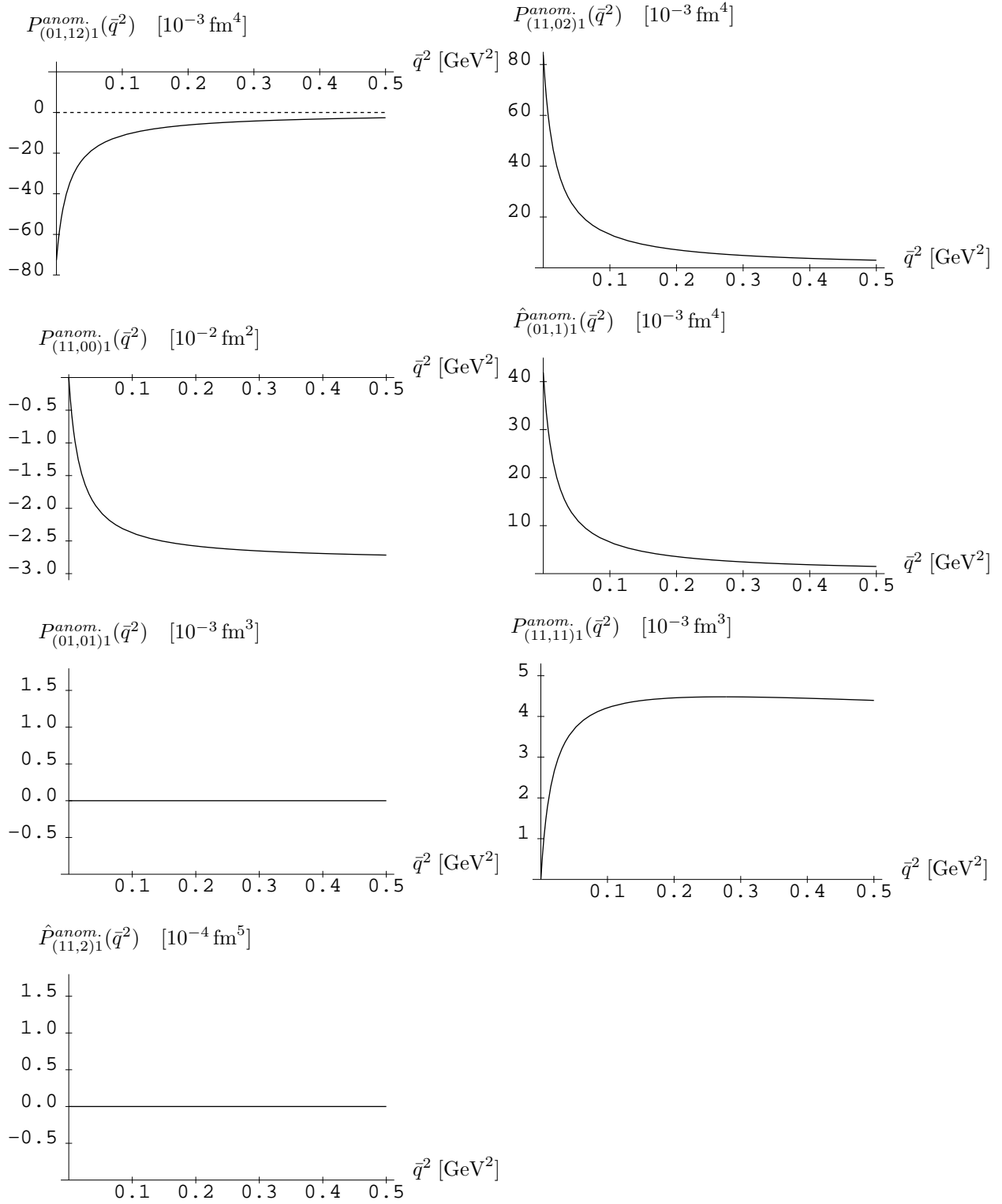}}
\vspace{1.5cm}
\caption[diag]{\label{figanom1} ${\cal O}(p^3)$ HBChPT/${\cal O}(\epsilon^3)$
SSE
$\pi^0$-pole  contributions to
the generalized spin polarizabilities.}
\end{figure}

\end{document}